\documentclass[11pt]{elsarticle}
\usepackage{epsfig,amsfonts,amssymb,amsmath}
\usepackage{color}
\usepackage{array}
\usepackage{multirow}

\makeatletter
\def\ps@pprintTitle{%
  \let\@oddhead\@empty
  \let\@evenhead\@empty
  \let\@oddfoot\@empty
  \let\@evenfoot\@oddfoot
}
\makeatother

\begin{document}

\begin{frontmatter}

\title{A network epidemic model with preventive rewiring:
\\ comparative analysis of the initial phase}

\author[address1]{Tom Britton}
\ead{tom.britton@math.su.se}

\author[address2]{David Juher}
\ead{david.juher@udg.edu}

\author[address2]{Joan Salda\~{n}a\corref{cor1}}
\ead{joan.saldana@udg.edu}
\cortext[cor1]{Corresponding author}

\address[address2]{Departament d'Inform\`{a}tica, Matem\`{a}tica Aplicada i Estad\'istica
\\
Universitat de Girona, Catalonia, Spain}

\address[address1]{Department of Mathematics,
\\
Stockholm University, Stockholm, Sweden}

\begin{abstract}
This paper is concerned with stochastic SIR and SEIR epidemic models on random networks in which individuals may rewire away from infected neighbors at some rate $\omega$ (and reconnect to non-infectious individuals with probability $\alpha$ or else simply drop the edge if $\alpha=0$), so-called preventive rewiring. The models are denoted SIR-$\omega$ and SEIR-$\omega$, and we focus attention on the early stages of an outbreak, where we derive expression for the basic reproduction number $R_0$ and the expected degree of the infectious nodes $E(D_I)$ using two different approximation approaches. The first approach approximates the early spread of an epidemic by a branching process, whereas the second one uses pair approximation. The expressions are compared with the corresponding empirical means obtained from stochastic simulations of SIR-$\omega$ and SEIR-$\omega$ epidemics on Poisson and scale-free networks. Without rewiring of exposed nodes, the two approaches predict the same epidemic threshold and the same $E(D_I)$ for both types of epidemics, the latter being very close to the mean degree obtained from simulated epidemics over Poisson networks. Above the epidemic threshold, pairwise models overestimate the value of $R_0$ computed from simulations, which turns out to be very close to the one predicted by the branching process approximation. When exposed individuals also rewire with $\alpha > 0$ (perhaps unaware of being infected), the two approaches give different epidemic thresholds, with the branching process approximation being more in agreement with simulations.
\end{abstract}

\begin{keyword}
network epidemic models,  preventive rewiring, branching process, pair approximation.
\end{keyword}

\end{frontmatter}

%\linenumbers

%%%%%%%%%%%%%%%%%%%%%%%%%%%%%%%%%%%%%%%%%%%%%%%%%%

\section{Introduction}\label{sec-intro}

Interactions among individuals in a population can be described by networks of who-contacts-whom. Studies of contact networks in sexually transmitted diseases have long revealed a high variability in the number of contacts per individual and highlighted the importance of those individuals described as "super-spreaders" for the onset of an epidemic \cite{Anderson, MHFM}. Similar conclusions about the importance of super-spread events were drawn from contact tracing data collected from recent epidemic outbreaks of airbone-transmitted diseases like those of the severe acute respiratory syndrome (SARS) in 2002 and 2003 \cite{Lipsitch, Riley}.

On the other hand, the risk perception among people during an epidemic outbreak triggers behavioural responses to lower the risk of contagion \cite{Lau,  Spring}, the avoidance of contacts with infected individuals being an example of such responses \cite{Fenichel}. This sort of social distancing led to the idea of disease-avoiding link rewiring and is one of the basis of the so-called adaptive or dynamic networks. Such a preventive rewiring assumes transmission of information which allows people to gather knowledge about the disease status of their neighbours. Therefore, in such networks the contact pattern is no longer static but evolves with the spread of an infectious disease according to the rules defining the rewiring process \cite{Gross06, JRS, Risau, Schwartz, Schwarzkopf, Zanette}.

Pairwise models have been the main approach adopted for the analysis of epidemic dynamics on adaptive networks \cite{Gross06, JRS, LJS, Risau, Schwarzkopf, TTK, Zanette}. This class of models was initially developed to deal with processes defined on regular (random) networks and offers a good description of their dynamics. In their classic formulation and over heterogeneous networks, however, their accuracy is far from being satisfactory, especially for its prediction of the epidemic threshold. The so-called effective degree models are extensions of them with a higher accuracy in their predictions (but also with a higher complexity). In these models, in addition to the disease status of nodes, the number of neighbours for each status is also considered \cite{House, Lindquist, Marceau}. At the individual level, pair-based epidemic models have been developed in terms of master equations for the probabilities of the individual pairs \cite{Frasca,Sharkey}. As with effective degree models, they show a higher accuracy than pairwise models formulated at the population level, but at the price of a higher computational complexity \cite{Sharkey}. 

To analyze epidemic outbreaks on dynamic networks, stochastic models have also been used. If the contact network has a large size and no cycles, it can be locally described as a tree and the initial phase of an epidemic can be approximated by a branching process \cite{DHB}. An example of a stochastic model defined on a dynamic network is the one developed in \cite{Volz07, Volz09}. The model assumes that, at a given rate, the identities of neighbours change stochastically by means of an instantaneous edge swap between a randomly selected pair of links. So, this neighbour-exchange mechanism is independent of the epidemic dynamics because it does not depend on the disease status of the involved nodes. In other words, it is not an example of behavioural response against the presence of the disease. Other models defined on dynamic networks whose architecture evolves by random edge swapping can be found in \cite{MSV}.

This paper aims mainly at comparing the predictions from both modelling methodologies (pairwise/stochastic) for the initial phase of Susceptible-Infectious-Recovered (SIR) and Susceptible-Exposed-Infectious-Recovered (SEIR) epidemics with preventive rewiring among individuals (so, with an interplay between the spread of the disease and the rewiring process, that is, between disease's dynamics and network dynamics). In particular, in the SIR model we will assume that susceptible individuals break off connections with infectious neighbors at a given rate $\omega$ and, in place of them, new connections to susceptible and recovered individuals are created with probability $\alpha$. As for the SEIR model, we consider two alternative scenarios for the dynamics of exposeds (i.e., infected but not infectious individuals). In the first one, exposed individuals break off with their infectious neighbors at a rate $\omega_{EI}$ and, with probability $\alpha$, they reconnect to any non-infectious individual in the population. In turn, susceptibles can also reconnect, with the same probability $\alpha$, to exposeds (in addition to other susceptibles and removed individuals) when breaking off with infectious neighbors. In the second scenario, exposed individuals do not rewire at all ($\omega_{EI}=0$), and susceptibles rewire away from both exposeds and infectives, and create new connections with probability $\alpha$. From a modeling viewpoint, the rewiring scenario can depend on whether exposed individuals realize they have been infected (for instance, because they show symptoms) or not (they are asymptomatic). In both scenarios, the degree distribution changes over time and its mean degree is preserved only when $\alpha=1$.

The introduction of a reconnection probability $\alpha$ allows us to consider different degrees of rewiring, ranging from a situation where each deleted link is replaced by a newly created one ($\alpha=1$) to the limit case where no new connection is made and edges are simply deleted ($\alpha=0$).  In other words, $\alpha$ can be though of a measure of the intensity of social distancing of rewiring individuals. As it is claimed in \cite{Fenichel}, people value person-to-person contacts and are willing to accept some disease risk to gain contact-related benefits. So, different values of $\alpha$ could be considered according to the type of social relationship modeled by the network.

The basic reproduction number $R_0$, namely, the average number of infections produced by a \textit{typical} infectious individual when the \emph{fraction} infected is still negligible, is one of the compared quantities. Its predicted value will be checked against stochastic simulations carried out on contact networks with degree distributions that follow a Poisson distribution and a power law, respectively. It is worth noting that, while the meaning of $R_0$ in randomly mixing homogeneous populations is straightforward because any infectious individual is as likely as any other to infect a susceptible one, in heterogeneous networks it requires that one specifies the meaning of typical individual \cite{DHM, Miller}. In most network models (in particular, for those without multiple levels of mixing), this definition implies that one has to compute $R_0$ from the average number of infections per infective once the early correlations of disease status around infectious individuals have been formed, which takes a couple of generations after the occurrence of the primary cases. Interestingly, this computation/redefinition of $R_0$ has been obtained under both previous modelling approaches \cite{Eames,Keeling99,PBT}. In fact, it is well known that both pairwise models and branching process approximations lead to the same epidemic (or invasion) threshold in networks without rewiring \cite{KG}.

In Section~\ref{sec-model} we present the SIR and SEIR epidemic models with rewiring, here denoted by SIR-$\omega$ and SEIR-$\omega$ respectively. In Section~\ref{SIR-branching} we use the branching process approximation to analyse the early phase of an SIR-$\omega$ epidemic. In particular, we compute $R_0$ and the expected degree of infectives during this initial phase as a function of the rewiring rate $\omega$. In Section~\ref{SEIR-branching} the same approximation is applied to the study of the early stage of an SEIR-$\omega$ epidemic under different types of rewiring processes. Section~\ref{Pairwise} contains the results of the initial phase obtained for both epidemic models using the pair approximation with the triple closure introduced in \cite{JRS,LJS} for heterogeneous networks. In particular, the SEIR-$\omega$ pairwise model extends the one considered in \cite{LJS} to account for the rewiring of exposed individuals and the possibility that susceptibles reconnect to exposed ones after breaking off an infectious link. In Section~\ref{simus} numerical estimates of $R_0$ and the mean degree of infectives during the initial phase are obtained from continuous-time stochastic simulations on heterogeneous networks. Finally, in Section~\ref{discuss} we discuss the analytical results obtained from both approximations and compare them to the output of the stochastic simulations. Moreover, we comment about the new insight into the role of the rewiring process in the SEIR-$\omega$ epidemic model.

\section{The stochastic network epidemic model with rewiring}\label{sec-model}

Let us define our stochastic network epidemic model. The population consists of a fixed number $N$ of individuals and the stochastic network model is given by the configuration model with degree distribution $D\sim \{ p_k\}$ having finite mean $\mu$ and finite variance $\sigma^2$ (e.g.\ \cite{Durrett}). This model is defined by all individuals having i.i.d.\ degrees $D_i$ and edge-stubs being pairwise connected completely at random with any loop or multiple edge being removed making the graph simple. We are primarily interested in the situation where $N$ is large, and the approximations will rely on this.

On this network we now define an epidemic model where susceptible individuals may rewire if they are neighbors of infectious individuals. We start by defining an SIR epidemic where infected people immediately become infectious and later recover, and then extend the model to an SEIR model in which infected people are at first exposed (latent), then they become infectious, and eventually they recover. The latter model is a bit more complicated in that now the rate of rewiring could differ depending on whether the person rewiring is susceptible or exposed, and depending on whether the person he/she rewires away from is exposed or infectious.

Note that such a rewiring process causes the degree distribution to change over time. However, since we will focus our analysis on the early stage of an epidemic outbreak where only a tiny fraction of individuals is initially infected, we will assume that the degree distribution $D$ does not change significantly during the initial phase of the epidemic.        

\subsection{The SIR-$\omega$ network epidemic with rewiring}
In this model individuals are at first susceptible. If an individual gets infected he/she immediately becomes infectious, and after some random time he/she recovers and becomes immune for the rest of the outbreak. SIR hence stands for susceptible-infectious-recovered (e.g.\ \cite{DHB} for more on SIR and SEIR epidemic models).

The SIR-$\omega$ model is defined on the network (described above) as follows.
Initially one randomly selected individual is infectious and the rest are susceptible. An infectious individual transmits the disease to each of its susceptible neighbors at a rate $\beta$, and the infectious periods are i.i.d.\ following an exponential distribution with rate parameter $\gamma$ (so infectious individuals recover at a rate $\gamma$). Further, susceptible individuals that are neighbors with infectious ones break off with such neighbors independently at rate $\omega$ and, with probability $\alpha$, replace each lost connection by reconnecting to a randomly selected non-infectious (i.e., susceptible or recovered) individual in the community. Therefore $\alpha \omega$ is the effective rewiring rate, i.e., the rate at which new links are created by susceptibles in substitution for those previously deleted.

The SIR-$\omega$ network epidemic has the following parameters: $\beta$ (infection rate), $\gamma$ (recovery rate), $\omega$ (rewiring rate), $\alpha$ (reconnection probability), and the degree distribution $D$ with mean $\mu$ and variance $\sigma^2$. We will focus on what happens early on in the epidemic, before a substantial fraction of the community has been infected.

\subsection{The SEIR-$\omega$ network epidemic with rewiring}
In the SEIR-$\omega$ model an infectious individual transmits the disease to each of its susceptible neighbors at a rate $\beta$, but, when this happens, the neighbor first becomes exposed (or latent) and can transmit the disease only after a time delay. In other words, such an exposed individual becomes infectious at a rate $\phi$, and at this moment can start infecting each of its susceptible neighbors at a rate $\beta$.
As before, infectious individuals recover at a rate $\gamma$.

As regards to rewiring, it can be modelled differently depending on when an individual starts and stops having a rewiring rate and also depending on which individuals it rewires away from. Our model considers three different rewiring rates: a susceptible individual rewires away from each exposed neighbor at a rate $\omega_{SE}$, a susceptible individual rewires away from each infectious neighbor at a rate $\omega_{SI}$, and an exposed individual rewires away from each infectious neighbor at a rate $\omega_{EI}$. It is of course possible to also allow for rewiring from other states, e.g.\ that susceptible individuals rewire from  recovered individuals, but since we are primarily focusing on the initial phase such rewiring would have negligible effects. If infected people are detected only when they become infectious (e.g.\ if the latent period is the same as the incubation period), and hence individuals rewire away from infected neighbors only when those become infectious and not while they are exposed, this would correspond to $\omega_{SE}=0$ and $\omega_{SI}=\omega_{EI}>0$: susceptible individuals are not aware of exposed (latent) neighbors being infected and hence do not rewire, and susceptible but also exposed (latent but unaware) individuals rewire away from the infectious ones. We could obtain a second scenario if also exposed (latent) individuals are known to have been infected (e.g.\ by contact tracing or because they show some symptoms). We would then have $\omega_{SE}=\omega_{SI}>0$ and $\omega_{EI}=0$: susceptible individuals rewire away both from exposed and infectious neighbors but exposed individuals do not rewire away since they know they have already been infected. In all cases, the reconnection probability $\alpha$ modulates the fraction of rewirings that are effectively done, and it is assumed to be the same both for susceptibles and for exposeds. Therefore, $\alpha \omega_{ij}$ is the effective rewiring rate of individuals in state $i$ away from individuals in state $i$.

The SEIR-$\omega$ network epidemic has all the parameters of the SIR-$\omega$ epidemic except that the rewiring rate $\omega$ now becomes three different rates: $\omega_{SE}$, $\omega_{SI}$ and $\omega_{EI}$, and there is a rate $\phi$ at which latent individuals become infectious.

\section{Branching process approximation of the initial phase of the SIR-$\omega$ epidemic}\label{SIR-branching}

Most stochastic epidemic models allow for a branching process approximation of the early stages of an outbreak, an approximation which can be made rigorous as the population size $N$ tends to infinity (e.g.\ \cite{BD95}). This applies also to network epidemics -- we now describe the approximation of the current model.

We derive expressions for the basic reproduction number $R_0$, here denoted by $R^{BA}_0$ to distinguish its expression from the one obtained using pair approximation. We also derive the exponential growth rate $r$ (the Malthusian parameter) for the situation that $R_0>1$, and the average degree of infected individuals. Since rewiring is a focus of this paper, we look at both the degree of newly infected individuals as well as on the average of all infectious individuals, the latter expected to be smaller than the former since individuals rewire away from infectious neighbors.

\subsection{The basic reproduction number $R_0$}

Recall that $R_0$ is defined as the mean number of new infections caused by a typical infected individual during the early stage of the epidemic. Individuals that get infected during the early stage will, at the time of infection, have the size biased degree distribution of neighbors, $\tilde D\sim \{\tilde p_k\}$, where $\tilde p_k= kp_k/\sum_jjp_j=kp_k/\mu$ (e.g.\ \cite{BJM-L}). One of the neighbors is its infector whereas the remaining neighbors, with large probability during the early stage of an outbreak, will be susceptible. When considering the disease progress it is only the $\tilde D-1$ susceptibles that are of interest since it is not possible to reinfect the infector.

From the derivation above, the mean number of susceptible neighbours a typical new infectee has during the early stages equals
$$
E(\tilde D-1)=\sum_k (k-1)\frac{k p_k}{\mu}=\mu-1+\frac{\sigma^2}{\mu}
$$
(e.g.\ \cite{BJM-L}). The probability to infect a given such neighbor is obtained by considering the competing events that may happen: there could be an infection (rate $\beta$), the neighbor could rewire away from the infectee  (rate $\omega$), or the infectee can recover (rate $\gamma$). The probability of infection hence equals $\beta/(\beta+\gamma+\omega)$. The basic reproduction number equals the mean degree multiplied by the transmission probability, i.e.
\begin{equation}
R_0^{BA}=\frac{\beta}{\beta+\gamma+\omega} E(\tilde D-1)=\frac{\beta}{\beta+\gamma+\omega} \left(\mu-1+\frac{\sigma^2}{\mu}\right) .\label{R_0}
\end{equation}
If there is no rewiring ($\omega=0$), the basic reproduction number equals $E(\tilde D-1)\beta/(\beta+\gamma)$ as is well known. Therefore, the rewiring reduces $R_0$, as expected. We note that $R_0^{BA}$ is independent of $\alpha$, so it has no effect on the beginning of an outbreak if rewired edges are dropped, always attached to new susceptible individuals or a mixture of two.

\subsection{The exponential growth rate $r$}

During the early stage and assuming a large population, the number of infectives in the epidemic will asymptotically (as $N\to\infty$) evolve like a branching process. $R_0$ is the corresponding mean offspring distribution. Another important quantity associated to this branching process is $\lambda(t)$, the expected birth rate (rate of new infections) of an infectee having ''age'' $t$, where age corresponds to time since infection. We now derive $\lambda(t)$ which in turn will help us derive the exponential growth rate of the epidemic.

As derived earlier, the average number of susceptible neighbors upon infection is $E(\tilde D-1)=\mu-1+\sigma^2/\mu$, and the infectee will infect each neighbor independently. The average rate of infection (=''birth'') for each neighbor is obtained by considering what must be fulfilled for infection to happen. In order to infect a neighbor $t$ time units after infection, the infectee must still be infectious, the neighbor should not have rewired, and the infectee should not yet have infected the neighbor. Given this, the infection rate equals $\beta$. Since all events are assumed to follow an exponential distribution this gives us the following expression for $\lambda (t)$:
\begin{equation}
\lambda(t)=E(\tilde D-1) \beta e^{-(\beta+\gamma+\omega)t}=\left(\mu-1+\frac{\sigma^2}{\mu}\right) \beta e^{-(\beta+\gamma+\omega)t} .
\end{equation}
The average total number of births (infections) is hence $\int_0^ \infty \lambda(t)dt=E(\tilde D-1)\beta/(\beta+\gamma+\omega)=R_0$  as it should be. The mean birth rate $\lambda (t)$ also determines the exponential growth rate of the epidemic, i.e.\ for which $r$, $I(t)\sim e^{rt}$ (cf. \cite{Jagers}). This $r$, the Malthusian parameter, is given by the solution of the Euler-Lotka equation
$$
\int_0^\infty e^{-rt}\lambda (t)dt =1.
$$
For our model this gives us, after a bit of algebra,
\begin{equation}
r=\beta E(\tilde D-2)- \gamma-\omega= \beta \left(\mu-2+\frac{\sigma^2}{\mu}\right) - \gamma-\omega .\label{r-SIR}
\end{equation}
Also the exponential growth rate is independent of $\alpha$.

\subsection{The mean degree of infectives}

We now turn to the mean degree of infectives during the early stages of the epidemic. We consider two different means. The first one is for newly infected, which in fact has already been shown: during the early stages newly infected individuals will have degree distribution $\tilde D$ (when considering the degree distribution we also count the non-susceptible infector), so the mean degree of newly infected individuals equals $E(\tilde D)=\mu +\sigma^2/\mu$.

The second mean, $E(D_I)$, denotes the average number of neighbors of \emph{all} infectives during the early stages, not only that of the newly infected ones. As described above, during the early stage of an outbreak the degree distribution of newly infected equals $\tilde D$. However, while still infectious, an individual loses susceptible neighbors by rewiring: each susceptible neighbor is lost at a rate $\omega$. The probability that a susceptible neighbor $v$ is still a neighbor (i.e.\ has not rewired) $t$ time units after our individual $x$ was infected and given that $x$ is still infectious, is obtained by conditioning on the potential infection time of the neighbor ($v$ only rewires if not yet infected). So, we have
\begin{align*}
P(\hbox{$v$ still a neighbor at }t) &=\int_0^t P(\hbox{$v$ still a neighbor at }s\, |\, \hbox{$v$ infected at }s)\beta e^{-\beta s}ds
\\
& \quad + P(\hbox{$v$ still a neighbor at }t\, |\, \hbox{$v$ not infected by  }t)e^{-\beta t}
\\
&= \int_0^t e^{-\omega s}\beta e^{-\beta s}ds  + e^{-(\beta+\omega)t}
\\
&= \frac{\beta}{\beta+\omega} + \frac{\omega}{\beta+\omega} e^{-(\beta+\omega)t} .
\end{align*}
Note that we condition on that $x$ remains infectious at $t$.

When deriving the degree distribution of all infectives during the early stage, we have to take into account both this decrease of degree with age, but also the fact that, in the exponential phase of the epidemic (recall that $I(t)\sim e^{rt}$), ''young'' infectives will be over-represented. The ratio of individuals infected $s$ time units ago over the number of individuals infected at present equals $e^{-rs}$ due to the exponential growth rate. And only a fraction $e^{-\gamma s}$ of them are still infectious at present. Consequently, the fraction of infectives that were infected $s$ time units ago or longer equals $e^{-(r+\gamma)s}$, so the age distribution of infectives is exponential with parameter $r+\gamma$ (this is the so-called \textit{stable age distribution} of this branching process, \cite{Jagers}).

The mean degree of all infectives during the early stage is obtained by conditioning on their age:
\begin{align}
E(D^{BA}_I) &=1+\int_0^\infty  E(\tilde D-1)\left(  \frac{\beta}{\beta+\omega} + \frac{\omega}{\beta+\omega} e^{-(\beta+\omega)s}\right) (r+\gamma)e^{-(r+\gamma)s} ds
\nonumber \\
&= 1+E(\tilde D-1)\left( \frac{\beta}{\beta+\omega} + \frac{\omega(r+\gamma)}{(\beta+\omega)(\beta+\omega+r+\gamma)} \right)
\nonumber \\
&= E(\tilde D) - \frac{\omega}{\beta},
\label{E(D_I) branching}
\end{align}
where, as before, $E(\tilde D)=\mu+\sigma^2/\mu$, and $r=\beta E(\tilde D-2) - \gamma-\omega $ was defined in Equation (\ref{r-SIR}). This mean degree should be valid after a couple of generations and will then change as the depletion of susceptibles will start affecting things.

As seen in (\ref{E(D_I) branching}) also the mean degree of infectives is independent of $\alpha$. At first this might seem surprising since the degree of infectious individuals are affected by rewiring. However, an infectious individual can loose edges (which reduces the degree) due to susceptible neighbours rewiring away from the infective, but it does not affect the degree of the infective whether these links are simply dropped or the rewiring susceptible connects to new individuals.

\section{Branching process approximation of the initial phase of the SEIR-$\omega$ model}\label{SEIR-branching}

We now study the extended SEIR model recalling that individuals who get infected are now first latent for an exponentially distributed time with rate parameter $\phi$, after which they become and remain infectious according to earlier rules. Individuals rewire away from infected neighbors. More precisely, a susceptible individual rewires from each exposed (latent) neighbor at a rate $\omega_{SE}$ and from each infectious neighbor at a rate $\omega_{SI}$. Moreover, exposed individuals rewire away from infectious neighbors at a rate $\omega_{EI}$. Of course, some of these rewiring intensities may be zero (cf. Sec 2.2). As before, upon each rewiring event the individual reconnects to a randomly chosen susceptible or recovered individual with probability $\alpha$ and with the remaining probability the edge is simply dropped.

\subsection{The basic reproduction number $R_0$ for the SEIR-$\omega$ model}

During the early stage of an outbreak, at the time of infection an exposed individual $e$ has degree distribution $\tilde D$ as before, one neighbor $i$ being the infector and the remaining neighbors being susceptible. So, $e$ has $E(\tilde D)-1$ expected susceptible neighbors. However, in the SEIR model this number can eventually increase by one, provided that $e$ can rewire away from its infector $i$ (if not yet recovered) and reconnect to a susceptible individual that will become a new neighbor. The probability for this to happen is $\alpha\omega_{EI}/(\phi+\omega_{EI})$. In consequence, the expected number of susceptible neighbors for $e$ is $E(\tilde D)-1+\alpha\omega_{EI}/(\phi+\omega_{EI})$. Any such neighbor, say $s$, will get infected if $e$ first becomes infectious (before $s$ rewires away from $e$) and, then, an infection occurs (before $e$ recovers or $s$ rewires from $e$). Hence, the probability for this to happen is the product of the probabilities of these two events, namely, $\phi/(\phi+\omega_{SE}) \cdot \beta/(\beta+\gamma+\omega_{SI})$. To conclude, the expected number of neighbors infected by $e$ equals:
\begin{equation}
R^{BA}_0 = \frac{\phi\beta}{(\phi+\omega_{SE})(\beta + \gamma+\omega_{SI})}\left( E(\tilde D)-1 + \frac{\alpha \, \omega_{EI}}{\phi+\omega_{EI}}\right) . \label{R0-SEIR}
\end{equation}

By studying Eq.~(\ref{R0-SEIR}) we make the following observations. If there is no rewiring from any state, $R^{BA}_0$ reduces to $\beta/(\beta+\gamma)\,E(\tilde D-1)$, i.e.\ the same as for the SIR case. Further, $R^{BA}_0$ is decreasing in both $\omega_{SE}$ and $\omega_{SI}$ as expected. However, $R^{BA}_0$ \emph{increases} with $\alpha$ and $\omega_{EI}$. If $\omega_{SE}=0$ and $\omega_{SI}=\omega_{EI}=\omega$, the perhaps most realistic example discussed in Sec. 2.2, then $R^{BA}_0$ can be increasing in $\omega$ for some parameter set-ups, implying that the quicker individuals rewire the larger epidemic outbreak! The explanation to this is that, when $\alpha\omega_{EI}>0$, the exposed (latent) individuals can rewire away from their infector to a susceptible neighbor, with the effect that they may later (once they become infectious) infect the new susceptible neighbor.

\subsection{The exponential growth rate $r$  for the SEIR-$\omega$ model}

As with the SIR-$\omega$ model, in order to compute the Malthusian parameter $r$ we first derive an expression for $\lambda (t)$, the average rate at which an individual, who was infected during the initial phase of the epidemic, infects new individuals $t$ time units after his/her time of infection.

At the time when an individual gets infected he has on average $E(\tilde D)$ neighbors, one being infectious (its infector) and the remaining $E(\tilde D-1)$ will, with large probability since we are in the beginning of the epidemic, be susceptible. At a rate $\beta$, the infected individual infects each of the $E(\tilde D-1)$ initially susceptible neighbors $t$ time units after infection if the following conditions are fulfilled: the infected individual must have terminated the latent period without the neighbor having rewired, and after that the infectious period should still be active, an infection should not yet have taken place, and the neighbor should not have rewired. As mentioned in the previous subsection, it is also possible that the infected individual infects through the link to the infector. This happens with a rate $\beta$ at $t$ time units after infection if the following holds: the infected individual rewired from its infector (and connects to a susceptible neighbor) while still latent, and after this the infected individual has become infectious, has not yet infected the neighbor nor has the new neighbor rewired. The above reasoning leads to the following expression for $\lambda (t)$:
\begin{align}
\lambda (t)&= \beta E(\tilde D-1) P(\text{infectious, neighbor did not rewire, neighbor not yet infected, at $t$}) \label{Lambda_SEIR}
\\
& + \beta P(\text{rewired while latent, infectious, neighbor not rewired, neighbor not infected, at $t$}).\nonumber
\end{align}
By conditioning on the end of the latency period, the first probability equals
$$
\int_0^t \phi \, e^{-\phi s}e^{-\omega_{SE}s} e^{-\gamma (t-s)}e^{-\omega_{SI}(t-s)} e^{-\beta(t-s)} ds.
$$
The second probability, now conditioning both on the time of first rewiring and the end of the latent period, equals
$$
\int_0^t\int_0^s \phi \, e^{-\phi s} \alpha\omega_{EI}e^{-\omega_{EI}u} e^{-\omega_{SE}(s-u)}e^{-\gamma (t-s)} e^{-\omega_{SI}(t-s)}e^{-\beta(t-s)} du \, ds.
$$
Using these expressions in Equation (\ref{Lambda_SEIR}) and solving the integrals results in the following expression for $\lambda(t)$:
\begin{align*}
 \lambda(t) &= E(\tilde D-1)\beta\phi \frac{e^{-(\omega_{SI}+\gamma+\beta)t}-e^{-(\phi+\omega_{SE})t}}{\phi+\omega_{SE}-(\omega_{SI}+\gamma+\beta)}
\\
&\qquad+\beta\frac{\phi \alpha\omega_{EI}}{\omega_{SE}-\omega_{EI}} \left( \frac{e^{-(\phi+\omega_{EI})t}-e^{-(\omega_{SI}+\beta+\gamma)t}}{\omega_{SI}+\beta+\gamma-(\phi+\omega_{EI})} - \frac{e^{-(\phi+\omega_{SE})t}-e^{-(\omega_{SI}+\beta+\gamma)t}}{\omega_{SI}+\beta+\gamma-(\phi+\omega_{SE})} \right) .
\end{align*}
We now use $\lambda (t)$ to derive the exponential growth rate $r$ of the epidemic in case it takes off, and also to confirm our expression for $R_0$. The latter is easy. If we compute $\int_0^\infty \lambda(t)dt$ using the expression above we get exactly $R_0$ as defined in Eq.\ (\ref{R0-SEIR}), as it should be. As for the Malthusian parameter $r$ this is given as the solution to the equation $\int_0^\infty e^{-rt}\lambda (t)dt=1$. For the expression of $\lambda(t)$ above, this can be shown to be equivalent to
\begin{equation}
\frac{\beta\phi}{(r+\omega_{SI}+\beta+\gamma)(r+\phi+\omega_{SE})} \left(E(\tilde D-1) +\frac{\alpha\omega_{EI}}{r+\phi+\omega_{EI}}\right) =1. \label{r-SEIR}
\end{equation}
For $\alpha\omega_{EI}>0$, this is a third order equation, but for positive values of $r$ (the relevant values as we assume $R_0>1$) the left hand side is decreasing in $r$, starting from a value larger than 1 when $r=0$ and decreasing to 0 as $r\to\infty$ implying that there is a unique solution to the equation.

Equation~(\ref{r-SEIR}) is not explicit but it is still possible to see how various parameters affect the growth rate. For example, $r$ is increasing in the infection rate $\beta$ and the mean degree $E(\tilde D)$. As regards to the rewiring rates, $r$ decreases in  $\omega_{SI}$ and $\omega_{SE}$ but increases with the "harmful" rewiring rate $\alpha\omega_{EI}$. Finally, as we increase the rate to leave the latent state (i.e.\ making the latent state shorter), the effect depends on other parameter values, but if we increase $\phi$ towards infinity it can be shown that we obtain the expression for $r$ of the SIR-$\omega$ model (cf.\ Eq.~(\ref{r-SIR})) as expected.

\subsection{The mean degree of infectives and related quantities }

For the SEIR-$\omega$ model it is also possible to derive specific features of infected individuals during the initial phase of an epidemic. For example, as we did for the SIR-$\omega$ case, we can compute $E(D_I)$, the average degree of infectives during the early stage of the outbreak. However, other mean quantities might be equally relevant as, for instance, the mean degree $E(D_L)$ of infected but still latent individuals, or the mean degree $E(D_{I+L})$ of either latent or infectives,  or for that matter the expected number of susceptible neighbors while in one of these states. For brevity and because some of these quantities have even more complicated expressions, we compute $E(D_I)$ without solving the integrals appearing in the derivation, and indicate how to modify the derivations if we want to compute another mean.

To compute $E(D_I)$, let us pick at random an infectious individual $i$ during the early stage of an outbreak and let $T$ denote how long ago this individual was infected. We first compute the expected degree of $i$ conditional upon $T=t$, denoted by $E(D_i|T=t)$. This is done by conditioning on the duration of the latent period $L=s$, which must lie between 0 and $t$ since $i$ is infectious $t$ time units after infection:
$$
E(D_i|T=t)=\int_0^tE(D_i|T=t,L=s)f_L(s|T=t)ds.
$$
The second factor is given by  $f_L(s|T=t)=\phi e^{-\phi s}e^{-\gamma (t-s)}/ \left( \int_0^t \phi e^{-\phi u}e^{-\gamma (t-u)}du\right)$. As for the first factor, the individual has $E(\tilde D)$ expected neighbours at the time of infection, one being infectious and the rest being susceptible. For the susceptible neighbors we compute the probability that they are still neighbors. For the infector, it could have lost a neighbor from this edge only if $i$ rewired away from the infector to a susceptible neighbor \emph{and} the new neighbor later rewired away from $i$. We hence get
\begin{align}
E(D_i|T=t,L=s) &=E(\tilde D-1)P(\text{susceptible neighbor did not rewire}|L=s, T=t) \label{cond-mean}
\\
&\hskip.5cm + 1-P(i\text{ looses edge to infector } |L=s, T=t).\nonumber
\end{align}
The first probability in (\ref{cond-mean}) is obtained by conditioning on whether the susceptible neighbor was infected or not, and, in the former case, whether the latent period ended before $t$ or not:
\begin{align*}
&P(\text{susceptible neighbor did not rewire} \, | \, L=s, T=t)
\\
&  = e^{-\omega_{SE}s} \left( \int_s^t \beta e^{-(\beta+\omega_{SI}) (u-s)}\left( \int_u^t \phi e^{-(\phi+\omega_{EI})(v-u)} dv +e^{-(\phi+\omega_{EI})(t-u)}\right) du + e^{-(\omega_{SI}+\beta)(t-s)}\right) .
\end{align*}
Here there should be no $\alpha$ in front of $\omega_{EI}$ because the event concerns an exposed neighbor and it is irrelevant for the degree of the infective whether this neighbor reconnects or not upon rewiring.

The second probability in (\ref{cond-mean}) consists of two events. Either $i$ rewired away from its infector and dropped the edge while exposed, or else $i$ rewired away from infector and reconnected while exposed, and the new neighbor rewired away from $i$. The first event has probability
$$
P(i\text{ rewired and dropped edge} \, | \, L=s, T=t) = (1-\alpha)\int_0^s \omega_{EI}e^{-(\omega_{EI}+\gamma) u} du.
$$
The second event is obtained by conditioning on the time when $i$ rewires away from the infector and reconnects to a susceptible neighbor, whether the second rewiring happens during the latency or infectious period of $i$, and in the latter case whether infection takes place or not:
\begin{align*}
&P( i\text{ rewired and reconnected, and new neighbor rewired away from }i \, | \, L=s, T=t)
\\
& = \int_0^s \alpha \omega_{EI}e^{-(\omega_{EI}+\gamma) u}   \left(  \int_u^s  \omega_{SE}e^{-\omega_{SE}(v-u)}dv \right) du +
\\
&
 \int_0^s \alpha\omega_{EI}e^{-(\omega_{EI}+\gamma) u}   \left(  e^{-\omega_{SE}(s-u)}  \int_s^t   e^{-(\beta+\omega_{SI})(v-s)} \left( \beta \int_v^t \omega_{EI} e^{-(\omega_{EI}+\phi)(z-v)}  dz+ \omega_{SI}  \right) dv\right) du .
\end{align*}
Note that $\alpha$ appears in front of $\omega_{EI}$ only when it concerns the infective, because then it has to reconnect to make spreading to a new individual possible, whereas there is no $\alpha$ when the rewiring refers to exposed neighbors rewiring away from the infective. In the latter case it is irrelevant for the degree of the infective whether or not the exposed individual reconnects upon rewiring.

It remains to derive the distribution for $T$, the time since infection. For this we know that due to the exponential growth rate $r$ of infectives, there is a fraction $e^{-rt}$ to choose from $t$ units earlier as compared to present time. However, we also require that the individual is infectious at present, an event which happens with probability $\int_0^t\phi e^{-\phi s}e^{-\gamma(t-s)}ds$. The probability density of $T$ is hence proportional to $e^{-rt}\int_0^t\phi e^{-\phi s}e^{-\gamma(t-s)}ds$, which after a bit of algebra gives the following density
$$
f_T(t)= \frac{(r+\gamma)(r+\phi)}{\phi-\gamma}\left( e^{-(r+\gamma)t}-e^{-(r+\phi)t}\right).
$$
Finally, the expected degree of a randomly chosen infective during the early stages is obtained by integrating with respect to this density:
\begin{equation}
E(D^{BA}_I)=\int_0^\infty E(D_i|T=t)f_T(t)dt.
\end{equation}
In order to compare this predicted $E(D_I)$ with the one obtained from pair approximation (see next section), for each set of values of the parameters
we obtain the value of $r$ given by the positive solution of \eqref{r-SEIR} and evaluate the resulting expression of the previous integral.

If we were to compute e.g.\ the average degree of a latent individual $E(D_L)$, we would similarly condition on the time $T$ since infection of the randomly chosen latent individual. This individual would have $E(\tilde D)$ neighbours at the time of infection, one infectious and the rest susceptible, and we need to compute the probability that these neighbours would not have been lost similarly to what we did before. We would then integrate this expected value with respect to the probability density of $T$, which is proportional to $e^{-rt}e^{-\phi t}$.

\section{$R_0$ and $E(D_I)$ for the SIR and SEIR pairwise models with rewiring}\label{Pairwise}

We now derive expressions for $R_0$ and $E(D_I)$ using an alternative deterministic approximation based on the closed-form equations for the dynamics of pairs of disease status, the so-called pairwise models. While the SIR-$\omega$ pairwise model was already introduced in \cite{LJS}, the SEIR-$\omega$ pairwise model is a generalization of the one also introduced in \cite{LJS} that includes rewiring of exposeds and reconnection rules introduced in accordance with it. This extended model will allow for a better understanding of the impact of rewiring on $R_0$ derived under this approach, here denoted by $R^{PA}_0$.

\subsection{The SIR-$\omega$ pairwise model}

Let $[S]$, $[I]$ and $[R]$ be the expected number of susceptible, infectious, and recovered individuals, respectively. Moreover, let $[ij]$ be the expected number of non-ordered connected \emph{$i$-$j$ pairs}, i.e. pairs whose individuals are in states $i$ and $j$, and let $[ijk]$ the expected number of connected non-oriented $i$-$j$-$k$ triples ($i,j \in \{S,I,R\}$). So, if $N$ is the network size and $L$ is the total number of links in the network, then $[S]+[I]+[R]=N$ and $[SS]+[SI]+[SR]+[II]+[IR]+[RR]=L$.

The SIR-$\omega$ model formulated in terms of triplets can be closed by assuming the statistical independence at the level of pairs which leads to the following approximations for the expected number of the involved triples:  $[SSI] \approx (E(\tilde{D}) -1) 2[SS][SI]/(\mu [S]) $, $[ISI] \approx 1/2 \cdot (E(\tilde{D})-1) [SI]^2/(\mu [S])$, and $[ISR] \approx (E(\tilde{D})-1) [SI][SR]/(\mu [S])$ with $\mu =2L/N$ being the average degree (see \cite{LJS} for details). Note that, since we focus our analysis on the early epidemic stage with a very small number of initially infectious nodes, we approximate the expected degree of the susceptible central node of a triple by $E(\tilde{D})$, the mean degree of a node reached by following a randomly chosen link in a wholly susceptible population at $t=0$, i.e., when the degree distribution is the initial one.

Upon introducing the triple closure into the original model, the initial dynamics of the SIR-$\omega$ model is determined by
\begin{eqnarray}
\frac{d}{dt}\,[S] &=& - \beta [SI], \nonumber \\
\frac{d}{dt}\,[I] &=& \beta [SI] - \gamma [I], \nonumber \\
\frac{d}{dt}\,[SI] &=& \left( \beta z \left( \frac{2[SS]}{[S]} - \frac{[SI]}{[S]} \right) - \beta - \gamma - \omega \right) [SI],  \nonumber \\
\frac{d}{dt}\,[II] &=& \beta \left( 1 + z \frac{[SI]}{[S]} \right) [SI] - 2 \gamma [II], \label{SIR-PA}\\
\frac{d}{dt}\,[SS] &=& \alpha \, \omega \frac{[S]}{N-[I]} [SI] -  \beta z \frac{2[SS]}{[S]} [SI],  \nonumber \\
\frac{d}{dt}\,[SR] &=& \left( \gamma + \alpha \, \omega \frac{[R]}{N-[I]} \right) [SI] - \beta z \frac{[IS]}{[S]}[SR], \nonumber \\
\frac{d}{dt}\,[IR] &=& 2 \gamma [II] - \gamma [IR] + \beta z \frac{[IS]}{[S]}[SR], \nonumber
\end{eqnarray}
where $z:=(E(\tilde{D}) - 1) / \mu$. Note that susceptible individuals break off (at a rate $\omega$) with infectious neighbours and reconnect (at a rate $\alpha \,\omega$) to other susceptibles or to recovered individuals with probability $[S]/(N-[I])$ and $[R]/(N-[I])$, respectively, in place of the deleted links.

The average number of susceptible individuals around an infective is $[SI]/[I]$. If this quantity stabilizes to a value $([SI]/[I])^*$ during the initial exponential growth of an epidemic, then we can use it to compute $R_0$ as (\cite{LJS})
\begin{equation}
R_0^{PA} := \frac{\beta}{\gamma} \left( \frac{[SI]}{[I]} \right)^* .
\label{R0_PA}
\end{equation}
Moreover, if $[II]/[I]$ and $[IR]/[I]$ also stabilize during this initial phase, then the mean degree of the infectives at this stage is given by
\begin{equation}
E(D^{PA}_I) = \left( \frac{[SI]}{[I]} \right)^* + \left( \frac{2[II]}{[I]} \right)^* + \left( \frac{[IR]}{[I]} \right)^*.
\label{ED_I_PA}
\end{equation}

The equations for the dynamics of the local densities $[SI]/[I]$, $2[II]/[I]$, and $[IR]/[I]$ are obtained from the SIR-$\omega$ model by using the standard rules of differentiation and are given by
\begin{eqnarray}
\frac{d}{dt} \left(\frac{[SI]}{[I]} \right) &=& - \left(\beta + \omega + \beta z \left( \frac{[SI]}{[S]} - \frac{2[SS]}{[S]} \right) +
\beta  \frac{[SI]}{[I]} \right) \frac{[SI]}{[I]} \, ,  \nonumber \\
\frac{d}{dt} \left(\frac{2[II]}{[I]} \right) &=& 2\beta \left( 1 + z \frac{[SI]}{[S]} \right) \frac{[SI]}{[I]}
- \left( \gamma + \beta \frac{[SI]}{[I]} \right) \frac{2[II]}{[I]} \, , \label{loc-densities}\\
\frac{d}{dt} \left(\frac{[IR]}{[I]} \right) &=& \gamma \frac{2[II]}{[I]} + \beta z \frac{[SI]}{[S]} \frac{[SR]}{[I]} - \beta \frac{[SI]}{[I]} \frac{[IR]}{[I]}. \nonumber
\end{eqnarray}
Now, taking the limit when $2[SS]/[S] \to \mu=2L/N$, and $[SI]/[S] \to 0$, i.e., after the introduction of the first infectious individuals, we obtain the following limit system:
\begin{eqnarray}
\frac{d}{dt} \left(\frac{[SI]}{[I]} \right) &=& \left(\beta (E(\tilde{D}) - 2) -  \beta  \frac{[SI]}{[I]} - \omega \right) \frac{[SI]}{[I]} \, ,
\nonumber \\
\frac{d}{dt} \left(\frac{2[II]}{[I]} \right) &=& 2\beta \, \frac{[SI]}{[I]} - \left(\gamma + \beta \frac{[SI]}{[I]} \right) \frac{2[II]}{[I]} \, ,
\label{limit-loc-densities} \\
\frac{d}{dt} \left(\frac{[IR]}{[I]} \right) &=& \gamma \frac{2[II]}{[I]} - \beta \frac{[SI]}{[I]} \frac{[IR]}{[I]} \, . \nonumber
\end{eqnarray}
It is interesting to observe that the reconnection probability $\alpha$ does not appear in the system. This means that disconnections from infectious individuals play a role in the early epidemic dynamics, but the way the new connections are created, or even if they occur at all ($\alpha=0$), does not play any role at this stage.

\begin{figure*}[hf]
\begin{center}
\includegraphics[scale=0.45]{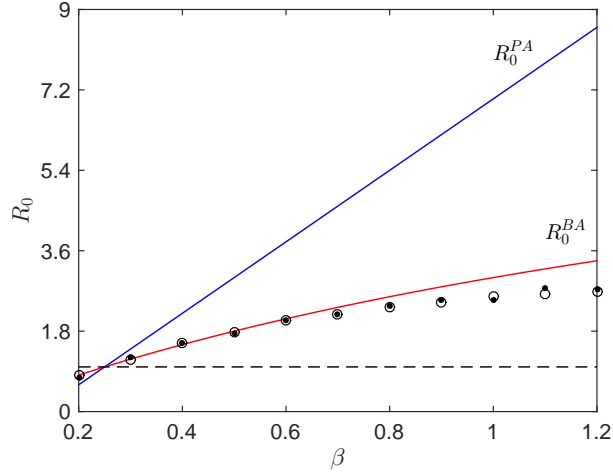}
\end{center}
\caption{Basic reproduction number of an SIR-$\omega$ epidemic as a function of the infection rate $\beta$. Dashed line corresponds to $R_0=1$. PA: pair approximation. BA: branching process approximation. Open circles (solid dots) correspond to $R_0$ computed from stochastic simulations of the epidemic on a Poisson (scale-free) network. Each network has a degree sequence with an average size-biased degree very close to $E(\tilde{D})=10$. This fact allows a direct comparison of outputs on each network. Parameters: $\gamma=1$, $\alpha=1$, $\omega=1$. \label{R0s} }
\end{figure*}

The first equation of \eqref{limit-loc-densities} is decoupled from the other two and has a unique positive equilibrium $([SI]/[I])^* =E(\tilde{D}) - 2 - \omega/\beta$, which is globally asymptotically stable. From this equilibrium and \eqref{R0_PA}, it immediately follows that
\begin{equation}
R_0^{PA} = \frac{\beta}{\gamma} \left( E(\tilde{D}) - \frac{\omega}{\beta} -2 \right)
\label{R0-SIR-PA}
\end{equation}
which defines the same epidemic threshold $R_0=1$ as $R^{BA}_0$ (cf.\ Eq.~\eqref{R_0}), but overestimates $R_0$ when it is larger than one. A graphical comparison of the expressions of $R_0$ obtained from each modelling approach is shown in Fig.~\ref{R0s} using $\beta$ as a tuning parameter.

The other components of the positive equilibrium of \eqref{limit-loc-densities} follow upon substituting $([SI]/[I])^*$ for $[SI]/[I]$ into the last two equations. Then, from \eqref{ED_I_PA}, we have
\begin{equation}
E(D^{PA}_I) = E(\tilde{D}) - \frac{\omega}{\beta},
\label{ED_SIR_PA}
\end{equation}
which is the same expression as the one obtained for $E(D_I^{BA})$ (cf. Eq.~\eqref{E(D_I) branching}).

\subsection{The SEIR-$\omega$ pairwise model}

According to the rewiring processes described in Section 2.2, and using the same notation and the triple closure as before, the equations of the SEIR model at the initial phase of the epidemic and with the three types of rewiring are
\begin{eqnarray}
\frac{d}{dt}[S] & = &-\beta [SI], \qquad  %\nonumber \\
\frac{d}{dt}[E] = \beta [SI]-\phi[E], \qquad % \nonumber \\
\frac{d}{dt}[I] = \phi [E]-\gamma [I], \nonumber \\
\frac{d}{dt}[SI] &=&\left(- \beta z \frac{[SI]}{[S]}-\beta -\gamma - \omega_{SI}\right)[SI] + \phi [SE],  \nonumber \\
\frac{d}{dt}[SE] &=& \beta z \left( \frac{2[SS]}{[S]}-\frac{[SE]}{[S]} \right) [SI] - \left(\phi+\omega_{SE}\right) [SE]
+ \alpha \, \omega_{EI} \frac{[S]}{N-[I]} [EI] + \alpha \, \omega_{SI} \frac{ \delta [E]}{N-[I]} [SI], \nonumber \\
\frac{d}{dt}[SS] &=&\left(-\beta z  \frac{2[SS]}{[S]} + \alpha \, \omega_{SI} \frac{[S]}{F([S],[E],[R])}\right)[SI]
+ \alpha \, \omega_{SE} \frac{[S]}{[S]+[R]} [SE],     \nonumber  \\
\frac{d}{dt}[SR] &=&\left(-\beta z \frac{[SR]}{[S]}+ \gamma + \alpha \, \omega_{SI} \frac{[R]}{F([S],[E],[R])}\right)[SI]
+\alpha \, \omega_{SE} \frac{[R]}{[S]+[R]} [SE],
\label{SEIR-PA} \\
\frac{d}{dt}[EI] &=& \beta \left(z \frac{[SI]}{[S]}+1\right)[SI]-\left(\phi + \gamma + \omega_{EI}\right)[EI]+2 \phi [EE], \nonumber \\
\frac{d}{dt}[EE] &=& \beta z \frac{[SE]}{[S]}[SI]-2 \phi [EE] + \alpha \, \omega_{EI} \frac{[E]}{N-[I]} [EI], \nonumber \\
\frac{d}{dt}[ER] &=& \beta z \frac{[SR]}{[S]}[SI] + \left( \gamma + \alpha \, \omega_{EI} \frac{[R]}{N-[I]} \right) [EI] - \phi [ER], \nonumber \\
\frac{d}{dt}[II]&=&\phi [EI]-2\gamma[II], \nonumber \\
\frac{d}{dt}[IR]&=& \phi [ER] + \gamma\left(2 [II]- [IR]\right), \nonumber
\end{eqnarray}
with $[S]+[E]+[I]+[R]=N$, $[SS]+[SE]+[SI]+[SR]+[EE]+[EI]+[ER]+[II]+[IR]+[RR]=L$, $z=(E(\tilde{D})-1)/\mu$, and $\mu=2L/N$ the mean degree of the initial degree distribution $D$. Moreover, $F([S],[E],[R])=[S]+[E]+[R]=N-[I]$ and $\delta=1$ if $\omega_{SE} = 0$, whereas $F([S],[E],[R])=[S]+[R]$ and $\delta=0$ otherwise. In the first case, susceptible individuals do not disconnect from exposed individuals and can reconnect to the latter when they rewire away from an infectious neighbor. In the second case ($\omega_{SE} >0$), susceptible individuals recognize exposed ones, rewire away from them, and only reconnect (with probability $\alpha$) to other susceptibles or to recovereds (so, $\delta=0$). Note that, in both cases, $F([S],[E],[R]) \to N$ at the early stage of an epidemic.

\begin{figure*}[hf]
\begin{tabular}{cc}
\hspace{-1cm}
\includegraphics[scale=0.45]{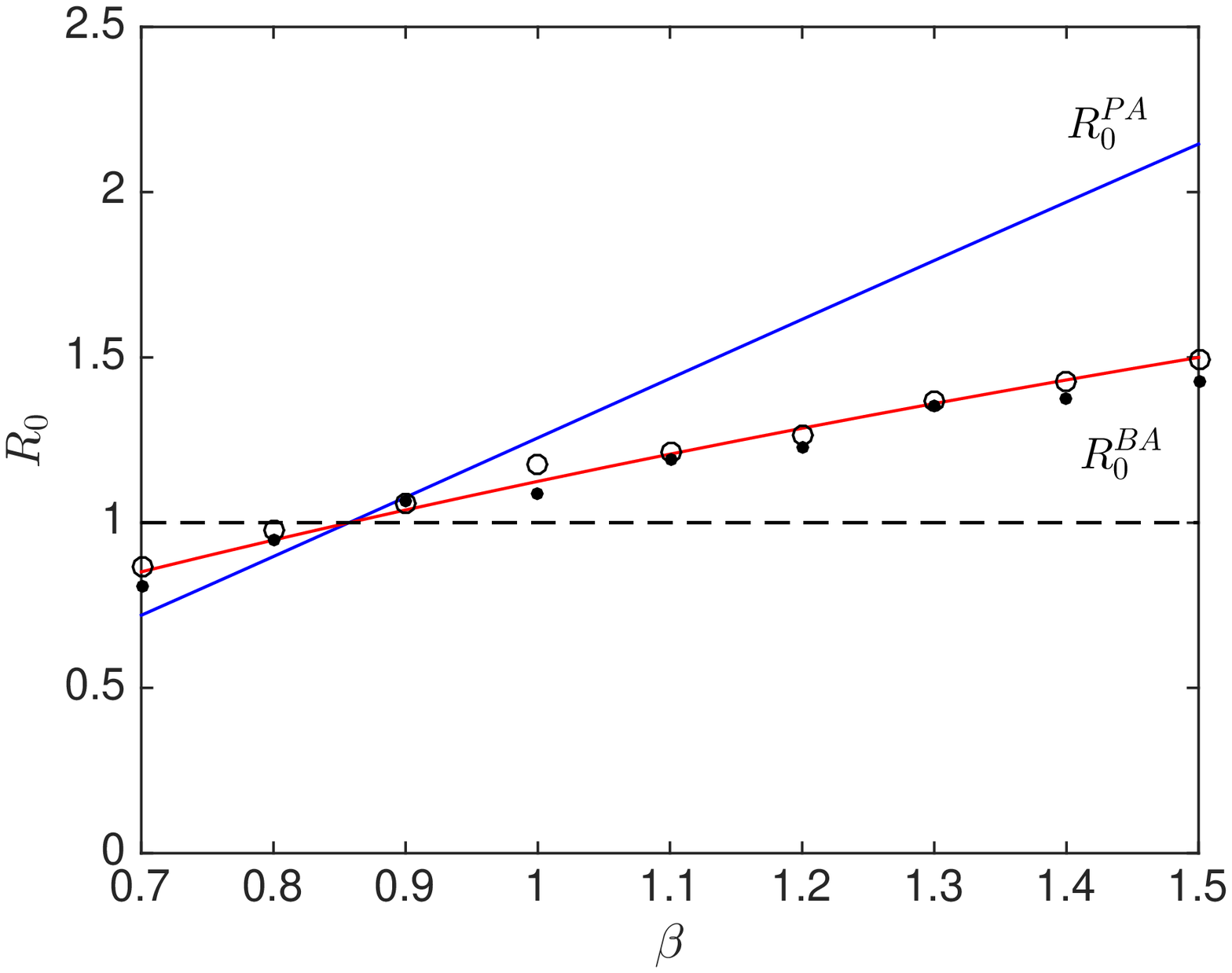}
&
\hspace{-1cm}
\includegraphics[scale=0.45]{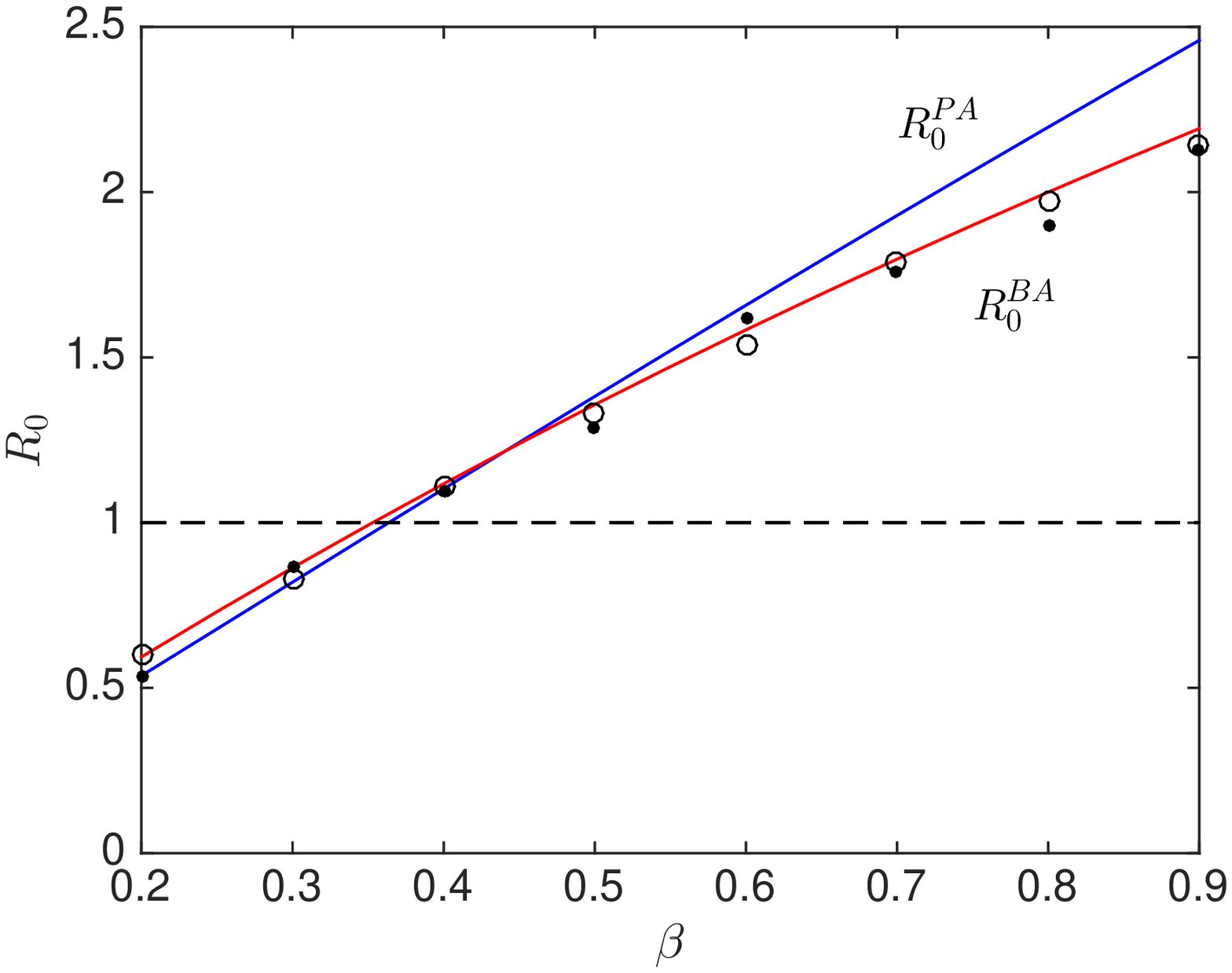}
\end{tabular}
\caption{Basic reproduction number, as a function of the infection rate $\beta$, of an SEIR-$\omega$ epidemic with $\omega_{SE}=1$, $\omega_{EI}=0$ (left) and $\omega_{SE}=0$, $\omega_{EI}=1$ (right). Dashed line corresponds to $R_0=1$. PA: pair approximation. BA: braching process approximation. Open circles (solid dots) correspond to $R_0$ computed from stochastic simulations of the epidemic on a Poisson (scale-free) network. Each network has a degree sequence with an average size-biased degree very close to $E(\tilde{D})=10$. Parameters: $\phi=1$, $\gamma = 2$, $\alpha=1$, and $\omega_{SI}=1$. Note that, when $\omega_{EI} > 0$, $R_0^{BA}$ is supercritical (i.e.\ larger than 1) for smaller $\beta$ than $R_0^{PA}$. \label{R0s2} }
\end{figure*}

If $\omega_{EI} > 0$, exposed individuals who break off a link with an infectious neighbor randomly reconnect to any susceptible, recovered, or exposed individual with a probability $\alpha [S]/(N-[I])$, $\alpha [R]/(N-[I])$, and $\alpha [E]/(N-[I])$, respectively. This corresponds to the situation where latent individuals are asymptomatic and, so, they do not know they have already got the infection. Therefore, one can also assume that susceptible individuals do not know the disease status of the exposed neighbours and take $\omega_{SE}=0$. Hence, susceptibles who break off a link with an infected neighbor reconnect to any susceptible, exposed or recovered individual with the same probabilities as the exposed ones, namely, $\alpha [x]/(N-[I])$ with $x \in \{[S],[E],[R]\}$. In Fig.~\ref{R0s2}, we show the predicted and observed $R_0$ as a function of the infection rate $\beta$. The values of the parameters are the same in both panels except for $\omega_{SE}$ and $\omega_{EI}$ which are 1 and 0 in the left panel, and 0 and 1 in the right one, respectively.

It is illustrative to check that the limit system for the dynamics of the local densities of disease status around an infectious individual (see Appendix) only contains one term with the reconnection probability $\alpha$, namely, the one with $\alpha \, \omega_{EI}$ as a prefactor. Therefore, the other contributions from the remaining rewiring rates in \eqref{SEIR-PA} do not appear when we restrict ourselves at the early stage of an epidemic. In other words, the precise rules for reconnecting susceptibles who have rewired away from an infectious neighbor, even if there is no reconnection at all of those individuals ($\alpha=0$), does not affect the epidemic dynamics during the initial phase. This claim, however, is not true for exposed individuals. The last term of the equation for $[SE]/[I]$ tells us that, when $\alpha \omega_{EI} > 0$, exposeds enhance the spread of the disease by rewiring away from infectious neighbors (because they will replace the latter with susceptible individuals) and, hence, $R^{PA}_0$ must increase under this rewiring.

The expression of $R_0$ defined by \eqref{R0_PA}, and computed from the corresponding positive equilibrium of the limit system for the local densities (see Appendix) is given by
\begin{equation}
R_0^{PA} = \frac{\phi}{\gamma} \, \xi^* \left( \xi^* - \frac{\gamma}{\phi} +1 \right)
\label{R0-SEIR-PA}
\end{equation}
with $\xi^*$ being the positive solution of Eq.~\eqref{xi} (in the Appendix) satisfying $\xi^* > \gamma/\phi -1$. If such a solution $\xi^*$ does not exist, then $([SI]/[I])^* =0$ is the only equilibrium value of $[SI]/[I]$ at the early stage of an epidemic and, hence, $R_0^{PA}=0$. From this expression, the one for $R^{BA}_0$, and Eq.~\eqref{xi} one obtains the following relationships between estimations of the epidemic thresholds (see Appendix for details):
\begin{itemize}
\item If $\alpha \, \omega_{EI} = 0$, then $\quad R_0^{PA} = 1 \quad \Longleftrightarrow \quad R^{BA}_0 = 1$.
\item If $\alpha \, \omega_{EI} > 0$, then $\quad R_0^{PA} = 1 \quad \Longrightarrow \quad R^{BA}_0 > 1$, \quad and $\quad R^{BA}_0 = 1 \quad \Longrightarrow \quad R_0^{PA} < 1$.
\end{itemize}
That is, both approaches predict the same epidemic threshold either when exposed individuals do not rewire ($\omega_{EI}=0$), or when there is no reconnection of broken links ($\alpha=0$). When the effective rewiring rate $\alpha \, \omega_{EI}$ is strictly positive, pair approximation predicts a higher epidemic threshold in terms of $\beta$ than the one obtained from the branching process approximation  (see the right panel in Fig.~\ref{R0s2}). Moreover, simulations show that $R_0^{PA}$ always overestimates the basic reproduction number when $R^{BA}_0 > 1$ if $\omega_{EI} = 0$. Finally, from Eq.~\eqref{xi} it follows that, for $\alpha>0$,  $\xi^*$ increases with $\omega_{EI}$, which implies that, as expected, $R^{PA}_0$ also increases with $\omega_{EI}$.

On the other hand, from the positive equilibrium of the limit system for the local densities, $E(D_I^{PA})$ is computed as $([SI]/[I])^*+(2[II]/[I])^*+([EI]/[I])^*+([IR]/[I])^*$ which gives
\begin{equation}
E(D_I^{PA}) = \left( (1+\xi^*) \phi - \gamma \right) \frac{\xi^*}{\beta} + \frac{((1+\xi^*) \phi - \gamma)(\xi^*+2) + \gamma)}{(1+\xi^*) \phi + \omega_{EI}}.
\label{E(D_I)PA}
\end{equation}
Note that, when there is no rewiring ($\omega_{SI}=\omega_{SE}=\omega_{EI}=0$), it follows that $E(D_I^{PA})=E(\tilde{D})$, as expected. Moreover, when $\omega_{EI}=0$, $\xi^*$ is given by \eqref{xi*} and $E(D_I^{PA})$ can be explicitly expressed in terms of the model parameters.

\begin{figure*}[hf]
\begin{tabular}{cc}
\hspace{-0.75cm}
\includegraphics[scale=0.42]{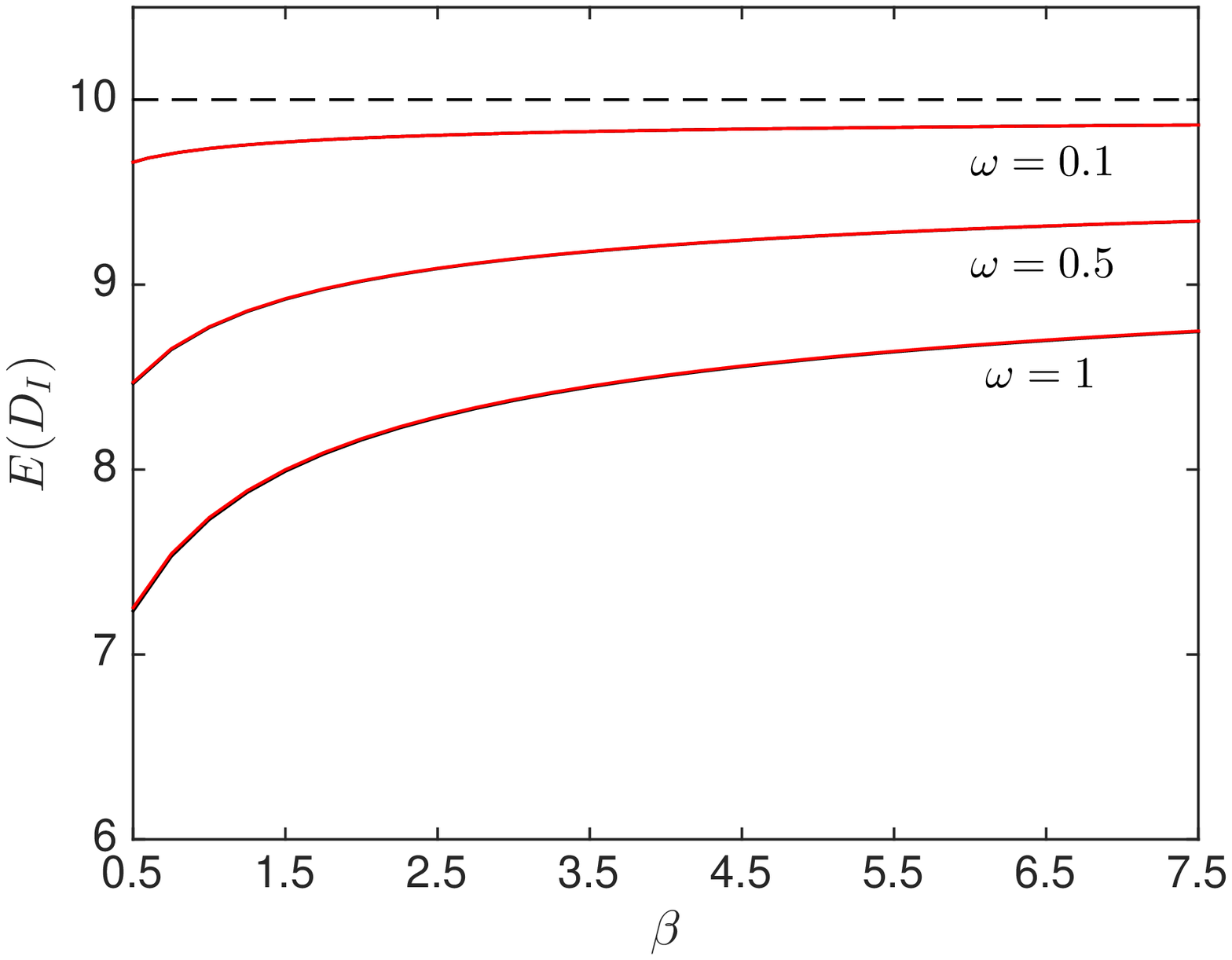}
&
\includegraphics[scale=0.42]{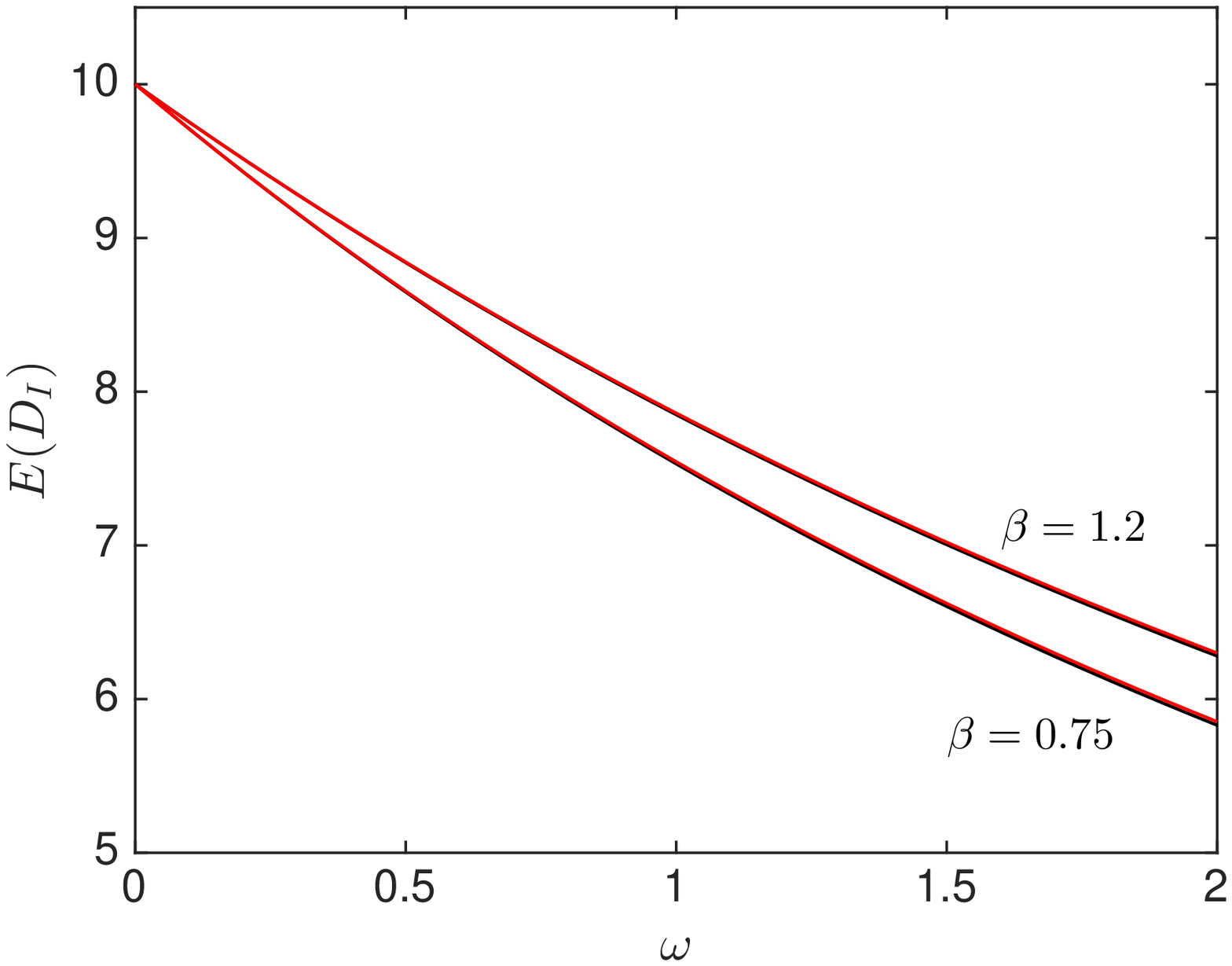}
\end{tabular}
\caption{Expected degree of the infectious individuals, $E(D_I)$, at the early stage of an SEIR-$\omega$ epidemic with $\phi=1$, $\gamma=2$, $\alpha=1$, $\omega_{SE}=0$, $\omega_{SI}=\omega_{EI}=\omega$, and $E(\tilde{D})=10$. Both predictions curves, $E(D^{BA}_I)$ and $E(D^{PA}_I)$, are almost graphically indistinguishable (the lower the thickness of the lines, the higher the overlap of the curves). Dashed horizontal line in the left panel corresponds to $E(D_I)$ without rewiring.}
\label{Comparison_E(Di)}
\end{figure*}

For $\alpha \, \omega_{EI}>0$ and $\omega_{SE}=0$, the numerical evaluation of the previous expression and its comparison with that of $E(D^{BA}_I)$  show that both predictions are very close to each other (see Fig.~\ref{Comparison_E(Di)}). Indeed, they are graphically distinguishable only for low values of $\beta$ (left panel) or for high values of the rewiring rate $\omega$ (right panel), i.e., for those parameters values that give $R_0$ close to 1. In these cases, differences occur at the second decimal place of the predicted mean degree. For $\omega_{EI}=0$ and $\alpha \in [0,1]$, both expressions give the same value for $E(D_I)$.

\section{Stochastic simulations}\label{simus}

To carry out continuous-time stochastic simulations we generated Poisson networks with $E(D)=9$ and scale-free (SF) networks with characteristic exponent 4 and minimum degree $k_{\min}=5$, i.e. $p(k)=3k^3_{\min} k^{-4}$. So, in both cases, $E(\tilde{D})=10$. All the networks had $N=10000$ nodes. The SF networks were generated using the configuration model algorithm. For each network and each combination of parameters, we averaged the outputs over 250 initial sets of 10 individuals infected uniformly at random (\emph{primary cases}). Moreover, we take the reconnection probability $\alpha=1$ in all simulations because it is when rewiring has the biggest effect. The stochastic time evolution of the infection spread was simulated by means of the Gillespie algorithm \cite{Gillespie}.

As mentioned in the introduction, since primary cases are selected at random regardless of their degree, a correct empirical computation of $R_0$ relies on counting the mean number of infections produced by the \emph{secondary cases} (individuals infected by the primary cases). So, for each experiment (that is, for each initial set of 10 random primary cases) we let the epidemic evolve until all primary and secondary cases have recovered. In Figs.~\ref{R0s}  and \ref{R0s2}, corresponding respectively to SIR-$\omega$ and SEIR-$\omega$ models, we compare the value of $R_0$ predicted by Eqs.~(\ref{R_0}) and (\ref{R0-SEIR}) with that obtained from Eqs.~(\ref{R0-SIR-PA}) and (\ref{R0-SEIR-PA}) respectively, and with the outputs of stochastic simulations carried out on a Poisson and a scale-free network. Since the variance of the theoretical SF degree distribution is quite high (it equals $3k_m^2/4=18.75$), there is a high variability among generated SF networks. Therefore, in order to compare the results for both types of networks in the same figure, we have chosen a random SF network whose degree sequence leads to a value of $E(\tilde{D})$ very close to the expected one ($\mu=7.5262$, $\sigma^2=18.6399$, and hence  $E(\tilde{D})=10.0029$).

\begin{figure*}[hf]
\begin{tabular}{cc}
\hspace{-0.75cm}
\includegraphics[scale=0.42]{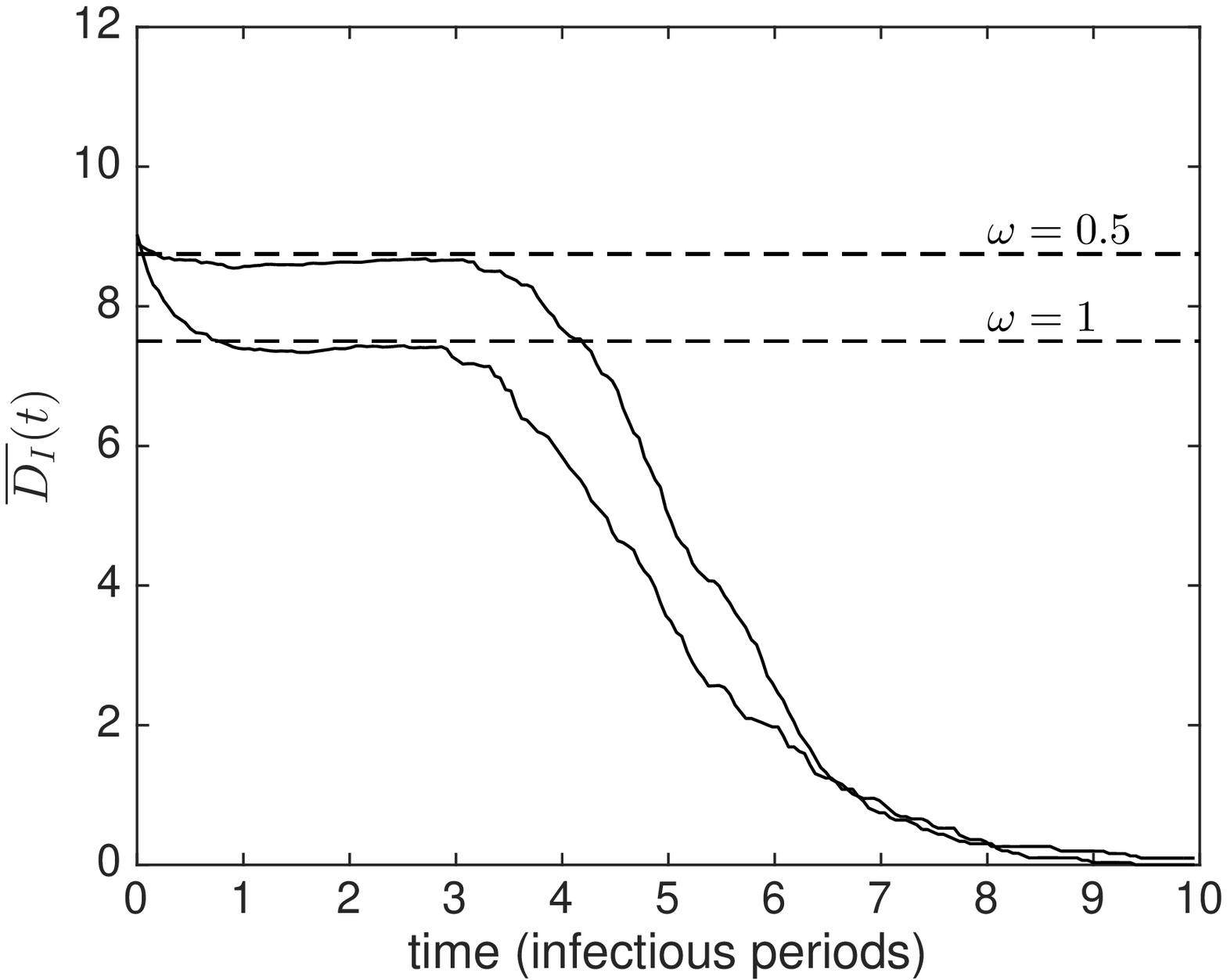}
&
\includegraphics[scale=0.42]{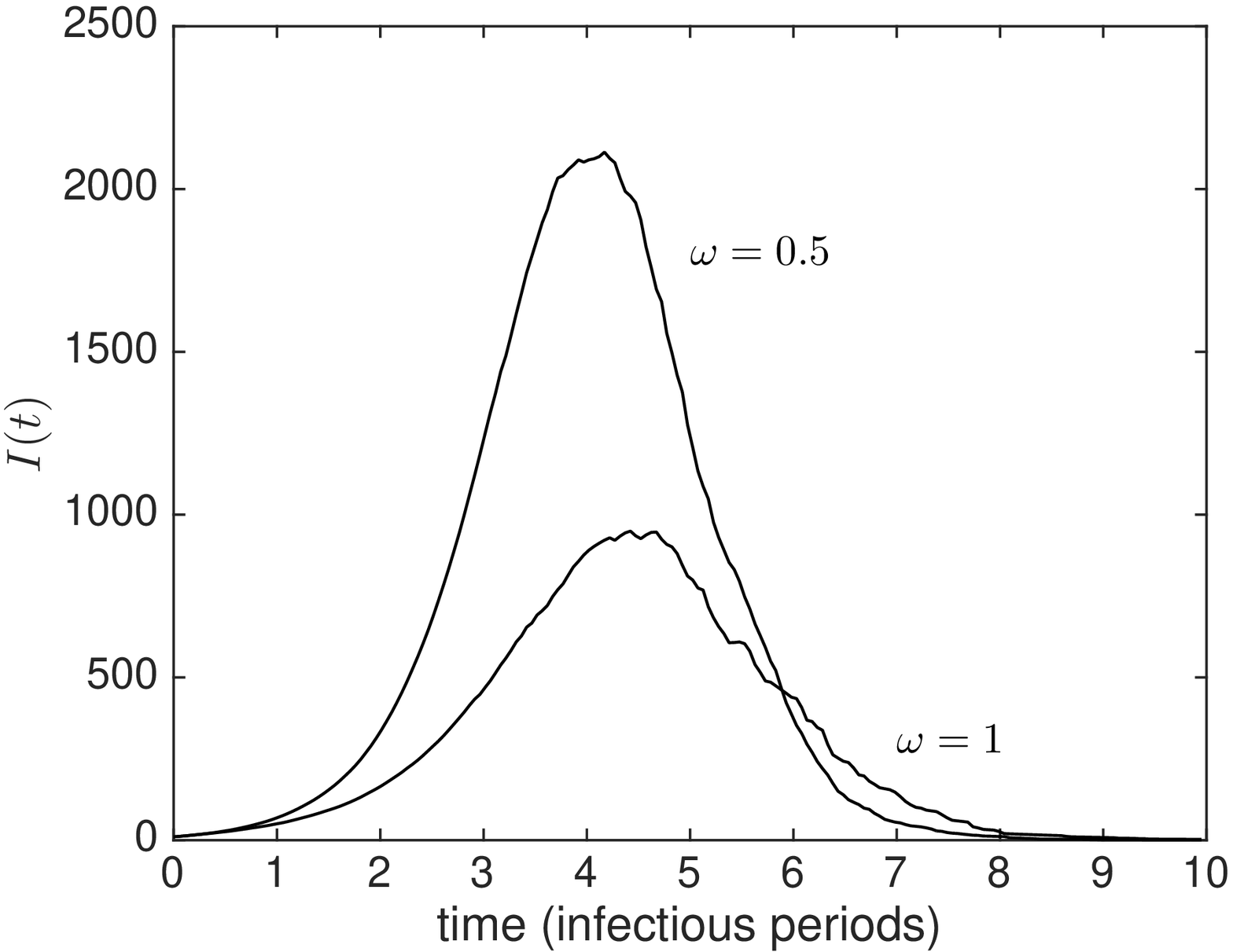}
\end{tabular}
\caption{Mean degree (left) and total number (right) of infectious nodes in an SIR-$\omega$ epidemic with $\gamma=1$ and $\beta=0.4$, over a Poisson network of mean degree $\mu=9$, and for $\alpha=1$ and $\omega\in\{0.5, 1\}$. Dashed horizontal lines correspond to the predicted $E(D_I)$ according to \eqref{E(D_I) branching}. Outputs averaged over 250 runs. Network size is $N=10000$ and the initial fraction of infected nodes is 0.1\%. The predicted mean degree during the exponential phase is close to the observed one.  \label{Poisson-SIR}}
\end{figure*}

\begin{figure*}[hf]
\begin{tabular}{cc}
\hspace{-0.75cm}
\includegraphics[scale=0.42]{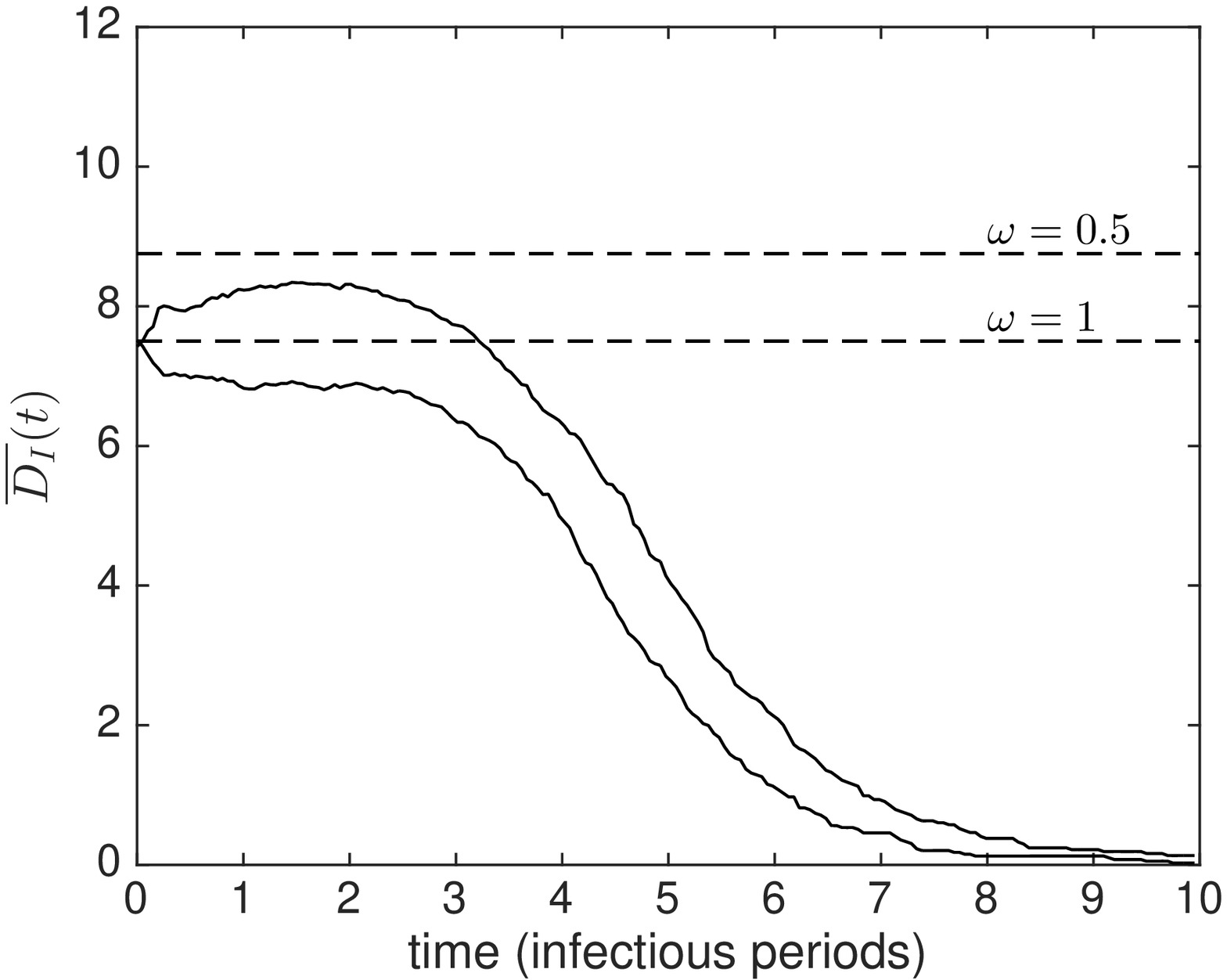}
&
\includegraphics[scale=0.42]{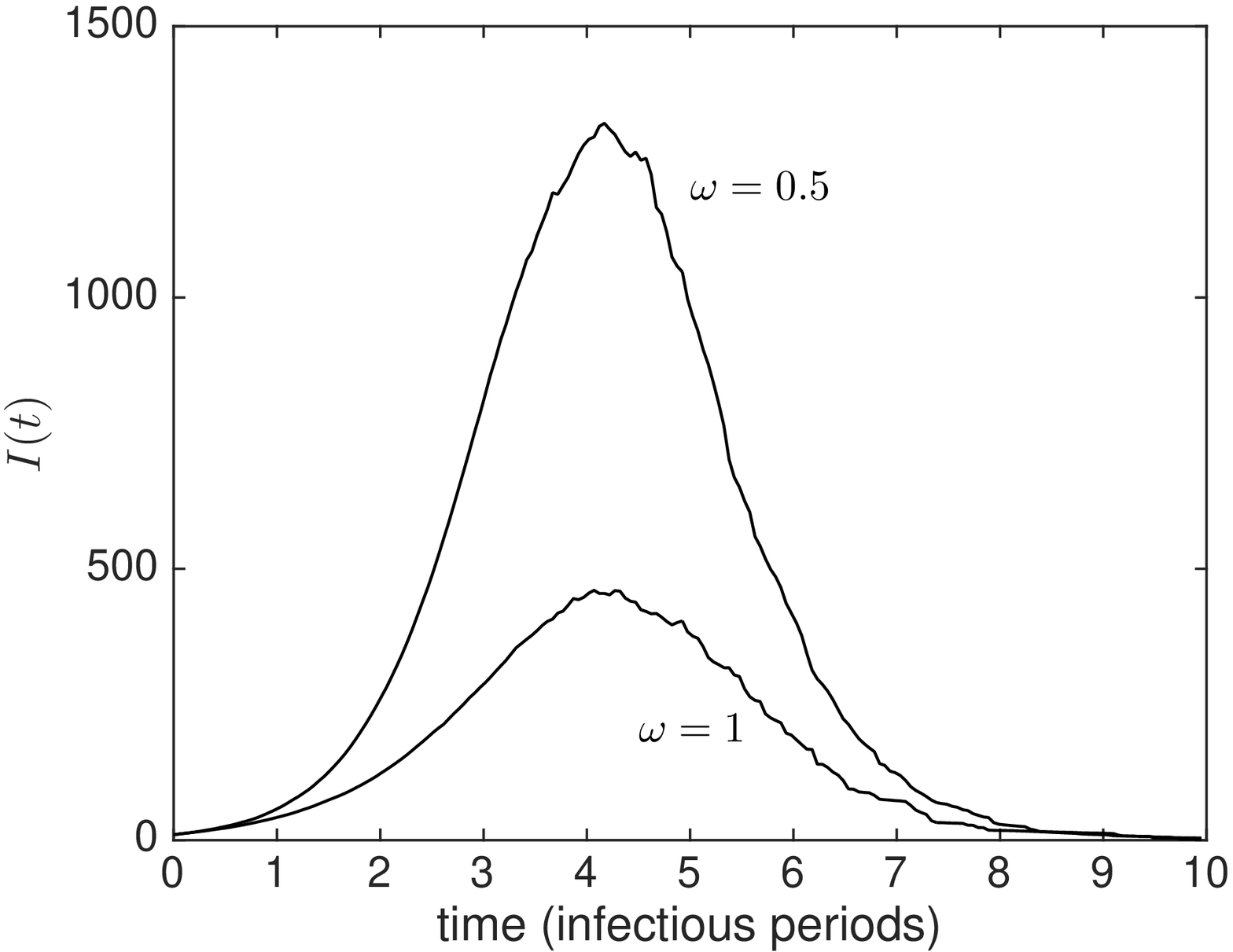}
\end{tabular}
\caption{Mean degree (left) and total number (right) of infectious nodes in an SIR-$\omega$ epidemic over a SF network with $p(k) \sim k^{-4}$ and $k_{\min}=5$ which gives a mean degree $\mu=3 k_{\min}/2 = 7.5$. Outputs averaged over 250 runs. Dashed horizontal lines correspond to  $E(D_I)$ according to \eqref{E(D_I) branching}. Parameters: $\gamma=1$, $\beta=0.4$, $\alpha=1$, $\omega\in\{0.5, 1\}$. $N=10000$ and the initial fraction of infected nodes is 0.1\%. \label{SF-SIR}}
\end{figure*}

We have also tested the accuracy of the analytical predictions for $E(D_I)$. Recall that, for the SIR-$\omega$ model, both approaches lead to the same value of $E(D_I)$ (cf.\ Eq.\ \eqref{E(D_I) branching} and \eqref{ED_SIR_PA}) whereas, for the SEIR-$\omega$ model, both predictions are very close to each other if $\alpha \, \omega_{EI}>0$, and are the same if $\alpha \, \omega_{EI}=0$. As it was mentioned, the predicted mean degree of the infectious individuals should be valid after a couple of generations, only as long as the growth of the epidemic is in its initial exponential phase, after which the depletion of susceptibles makes the hypothesis of the derivation no longer valid. In Figs.~\ref{Poisson-SIR}--\ref{SF-SEIR} we show, for two values of $\omega$, the evolution of the total number $I(t)$ of infectious individuals at time $t$, the average degree $\overline{D_I}(t)$ of the infected individuals, and the corresponding analytical predictions. At any given time $t$, the value of $\overline{D_I}(t)$ is computed as the total number of links containing an infected individual (the edges joining two infected are counted twice) over $I(t)$. The right panels of these figures show that the curve $I(t)$ fits to an exponential function (initial phase) on an interval $[0,t_e]$ with $t_e$ less than the time $I(t)$ attains its maximum. It is precisely on this interval that the mean degree $\overline{D_I}(t)$ on Poisson networks keeps almost stationary around a value close to the predicted one (see left panels of Fig.~\ref{Poisson-SIR} and ~\ref{Poisson-SEIR}). Such a plateau in the profile of $\overline{D_I}(t)$ is not so nicely observed when simulations are carried out on scale-free networks.

\section{Discussion}\label{discuss}

It is known that pairwise models for the spread of SIR-type diseases through static homogeneous networks predict the same epidemic threshold as the one obtained from the probabilistic computation of $R_0$ when infectious periods are exponentially distributed \cite{Keeling99,KG}. By using a pairwise model with a triple closure introduced in \cite{JRS,LJS}, and a branching process approximation of a stochastic network epidemic, we have seen that the same epidemic threshold is also predicted for dynamic networks whose topology evolves according to the preventive rewiring of susceptible individuals. As expected from a preventive rewiring, the higher the rewiring rate $\omega$, the lower $R_0$ is for both predictions, and this is true regardless of the value of the reconnection probability $\alpha$. However, for any $\omega$, the pair approximation overestimates $R_0$ when it is larger than 1 as compared to stochastic simulations and to the value $R_0^{BA}$ obtained from the branching process approximation. The reason is that the value $R_0^{PA}$ predicted by the pairwise model is a linear function of the infection rate $\beta$  and, hence, an unbounded number of new cases is predicted as $\beta$ increases (cf. Eq.~\eqref{R0-SIR-PA}). Such a linear dependence on the infection rate is a common feature of deterministic epidemic models \cite{Anderson}. In contrast, the hyperbolic dependence of $R_0^{BA}$ on $\beta$ (cf. Eq.~\eqref{R_0}) reflects the saturation in the production of new infections for high infection rates, and leads to values of $R_0$ that are closer to those obtained from the simulations.

The same relationships between estimates of $R_0$, and between epidemic thresholds, also hold for SEIR-$\omega$ models when susceptible (but not exposed/latent) individuals break off connections with their infectious/exposed neighbours, and reconnect to randomly chosen susceptible or recovered individuals with a given probability $\alpha$ ($\omega_{EI}=0$). However, if exposed individuals also disconnect from infectious neighbours and reconnect to randomly chosen non-infectious individuals ($\alpha \, \omega_{EI} > 0$), then the epidemic thresholds from the two approaches differ from each other, with $R_0^{BA} >1$ when $R_0^{PA}=1$. Interestingly, as long as $\alpha>0$, this rewiring of exposeds is not preventive but harmful since it does not help to contain the disease: sooner or later exposed individuals will become infectious and, when an exposed replaces its infector with a susceptible individual, the number of infections he/she can produce increases. This is why $R_0$ increases with $\alpha\omega_{EI}$, in contrast to what happens with the other two rewiring rates, $\omega_{SI}$ and $\omega_{SE}$.

Note that, because we are concerned with the initial stage of an epidemic, in large networks rewired links will, with a very high probability, point to susceptible individuals. Therefore, as long as only susceptible individuals rewire, the initial propagation of the disease will not be particularly affected by the type and intensity ($\alpha$) of the reconnection process. In fact, this is what follows from the computation of $R_0$ under both approaches. For the SIR-$\omega$ model, for instance, this can be easily seen from the derivation itself of $R_0$ under the branching process approximation since the former does not depend on how new connections (if any) are made. Similarly, from direct inspection of the limit system governing the dynamics of the local densities around infectious individuals (cf. Eq.~\eqref{limit-loc-densities}), one sees that there is no term corresponding to reconnection of links (i.e., terms with $\alpha \omega$ as a prefactor).

Both estimates of $R_0$ have been checked by obtaining, from stochastic simulations carried out on random networks, the mean number of infections produced not by the first infectious individuals landing in the population (primary cases), but by the second generation of infectives (secondary cases). It has been recognized elsewhere (\cite{Eames,Keeling99,PBT}) that this ``redefinition'' of $R_0$ for epidemics on networks is the suitable one because it takes into account the local correlations of disease status developed around infectives during the epidemic exponential growth (initial phase). Simulation results clearly indicate that the estimation of $R_0$ obtained from the branching process approximation is much better than the one derived from pairwise models, and gives the correct epidemic threshold when $\alpha \omega_{EI} > 0$. In particular, for the SEIR-$\omega$ model there is an excellent agreement for all the shown values of $\beta$ (Fig.~\ref{R0s2}), whereas for the SIR-$\omega$ model the agreement is not as good when $\beta$ is not close to its critical value (Fig.~\ref{R0s}).

On the other hand, for the SIR-$\omega$ model we have also seen that both approaches predict the same expected degree $E(D_I)$ of infectives at the early stage of an epidemic. In particular, it follows that $E(D_I)$ is a linear decreasing function of the rewiring rate $\omega$. Its computation from stochastic simulations clearly shows that, for moderately large values of $\omega$ and $\beta$, the mean degree of the infectious nodes $\overline{D_I}(t)$ remains quite constant during the exponential phase of the disease. Moreover, the agreement between theoretical predictions and observations using both $\omega$ and $\beta$ as tunable parameters is very good in Poisson networks for low values of the rewiring rates and moderate values of $\beta$. For high values of $\beta$, the exponential phase is so fast that the time window where $\overline{D_I}(t)$ is roughly constant is hardly noticeable. Similarly, when rewiring is high, $\overline{D_I}(t)$ decreases monotonously without any plateau during this initial phase. For scale-free networks and moderate values of $\beta$ and $\omega$, however, the predicted $E(D_I)$ overestimates the observed $\overline{D_I}(t)$ (cf. Fig.~\ref{Poisson-SIR} and Fig.~\ref{SF-SIR}).

\begin{figure*}[hf]
\begin{tabular}{cc}
\hspace{-0.75cm}
\includegraphics[scale=0.42]{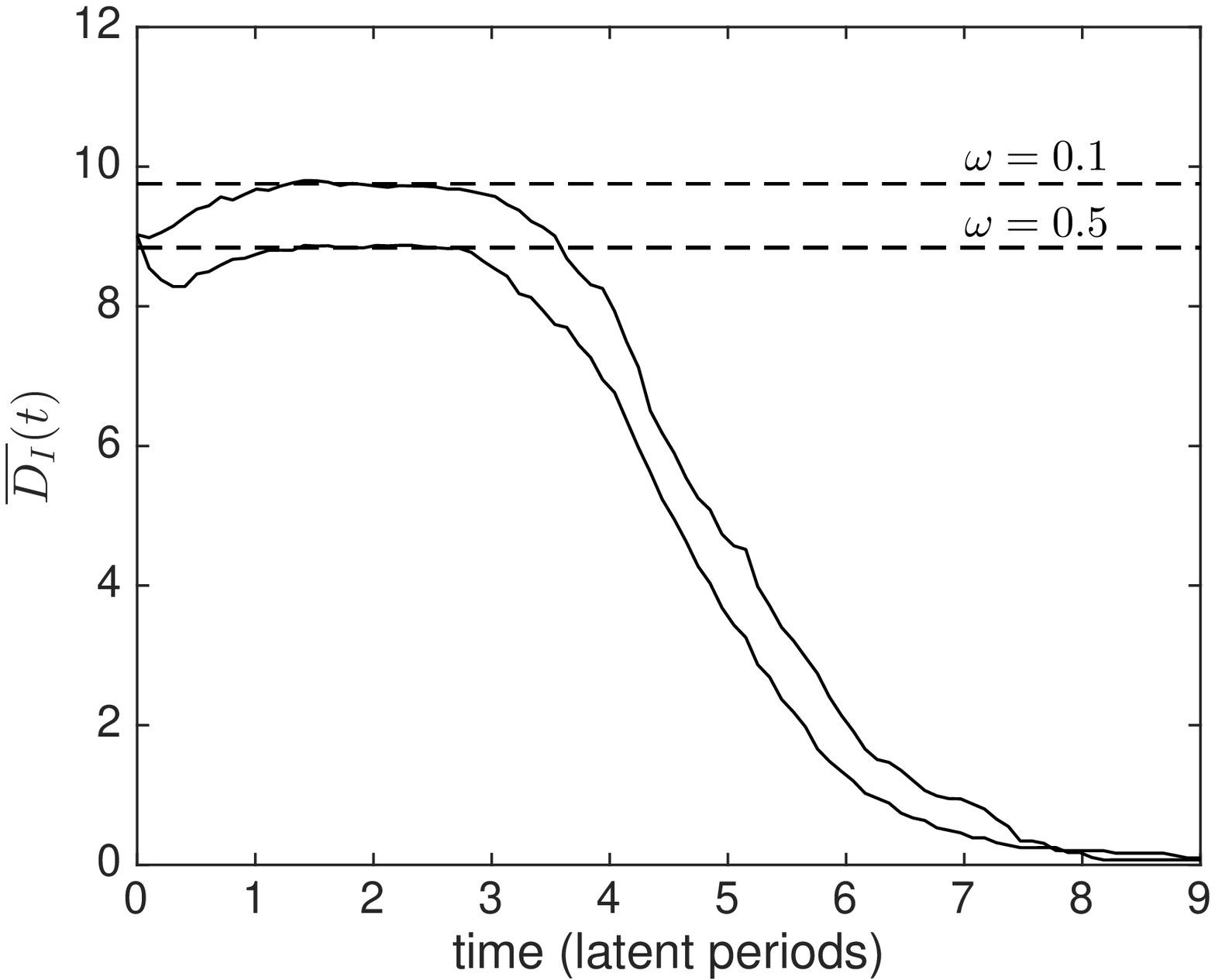}
&
\includegraphics[scale=0.42]{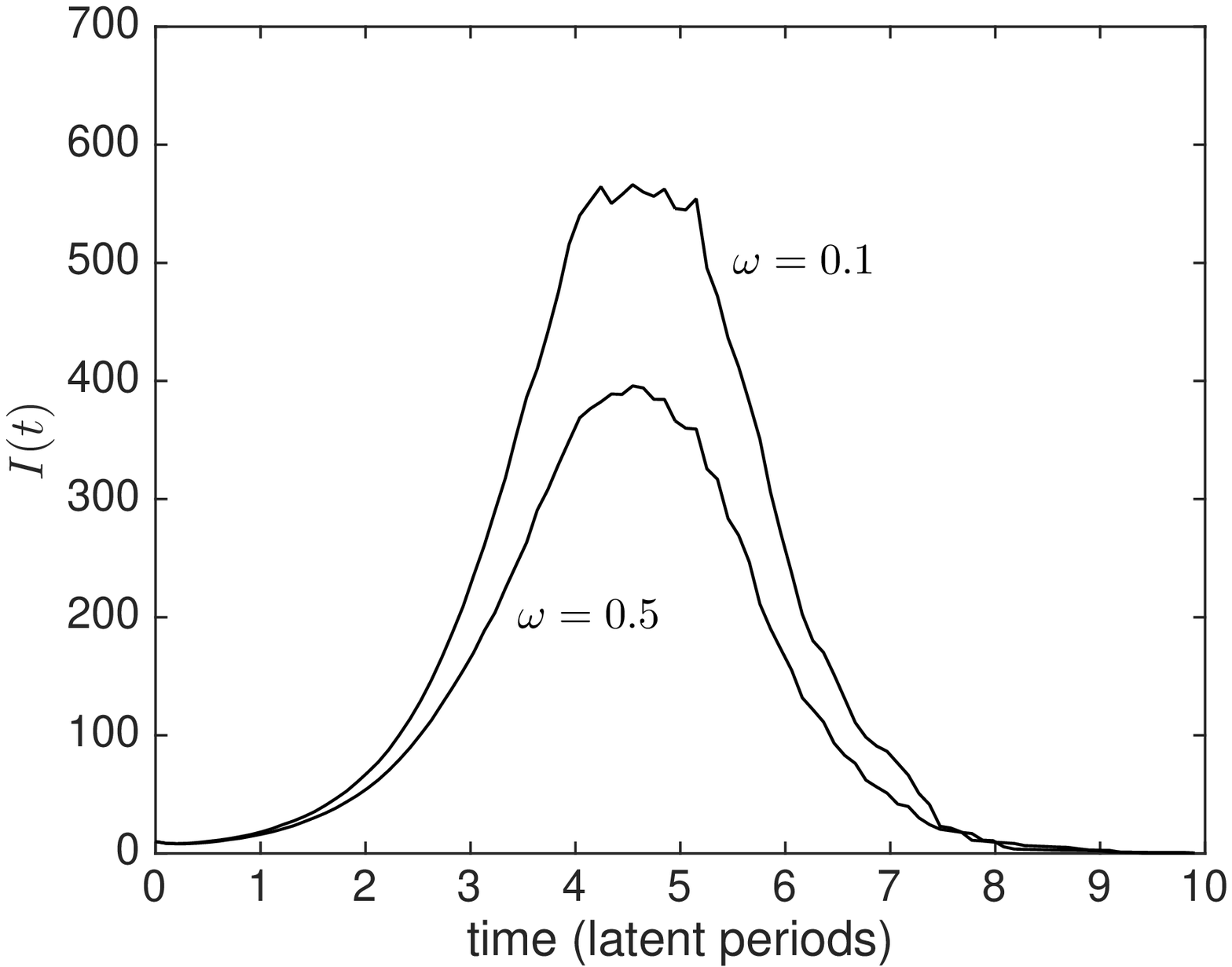}
\end{tabular}
\caption{Mean degree (left) and total number (right) of infectious nodes in an SEIR-$\omega$ epidemic over a Poisson network with mean degree $\mu=9$. Outputs averaged over 250 runs. Dashed horizontal lines correspond to the predicted $E(D_I)$: $9.75$ for $\omega=0.1$, and $8.84$ for $\omega=0.5$ (both approaches lead to these rounded values). Parameters: $\phi=1$, $\gamma=2$, $\beta=1.2$, $\alpha=1$, $\omega_{SE}=0$, and $\omega_{SI}=\omega_{EI}=\omega\in\{0.1, 0.5\}$. $N=10000$ and the initial fraction of infected nodes is 0.1\%. \label{Poisson-SEIR}}
\end{figure*}

\begin{figure*}[hf]
\begin{tabular}{cc}
\hspace{-0.75cm}
\includegraphics[scale=0.42]{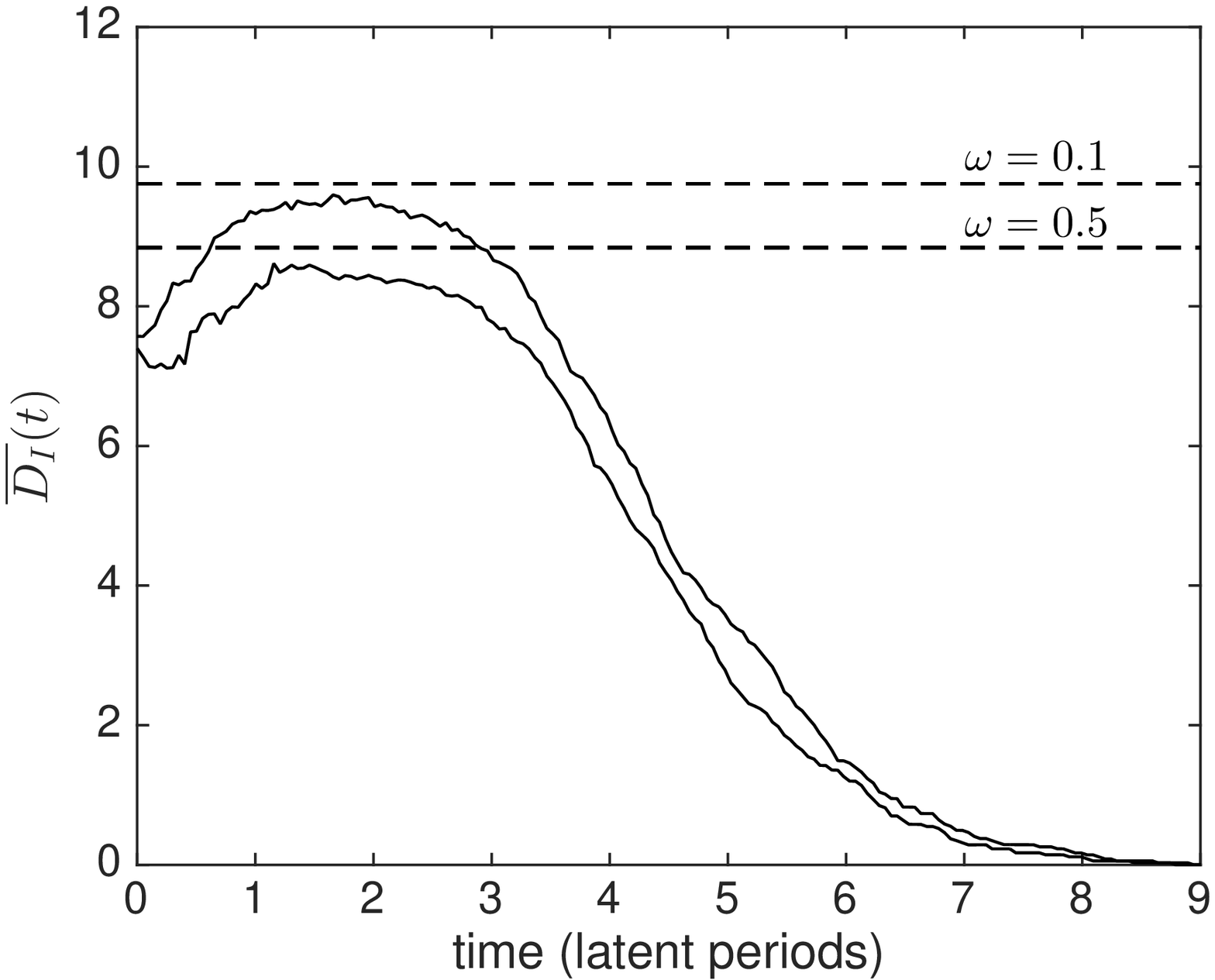}
&
\includegraphics[scale=0.42]{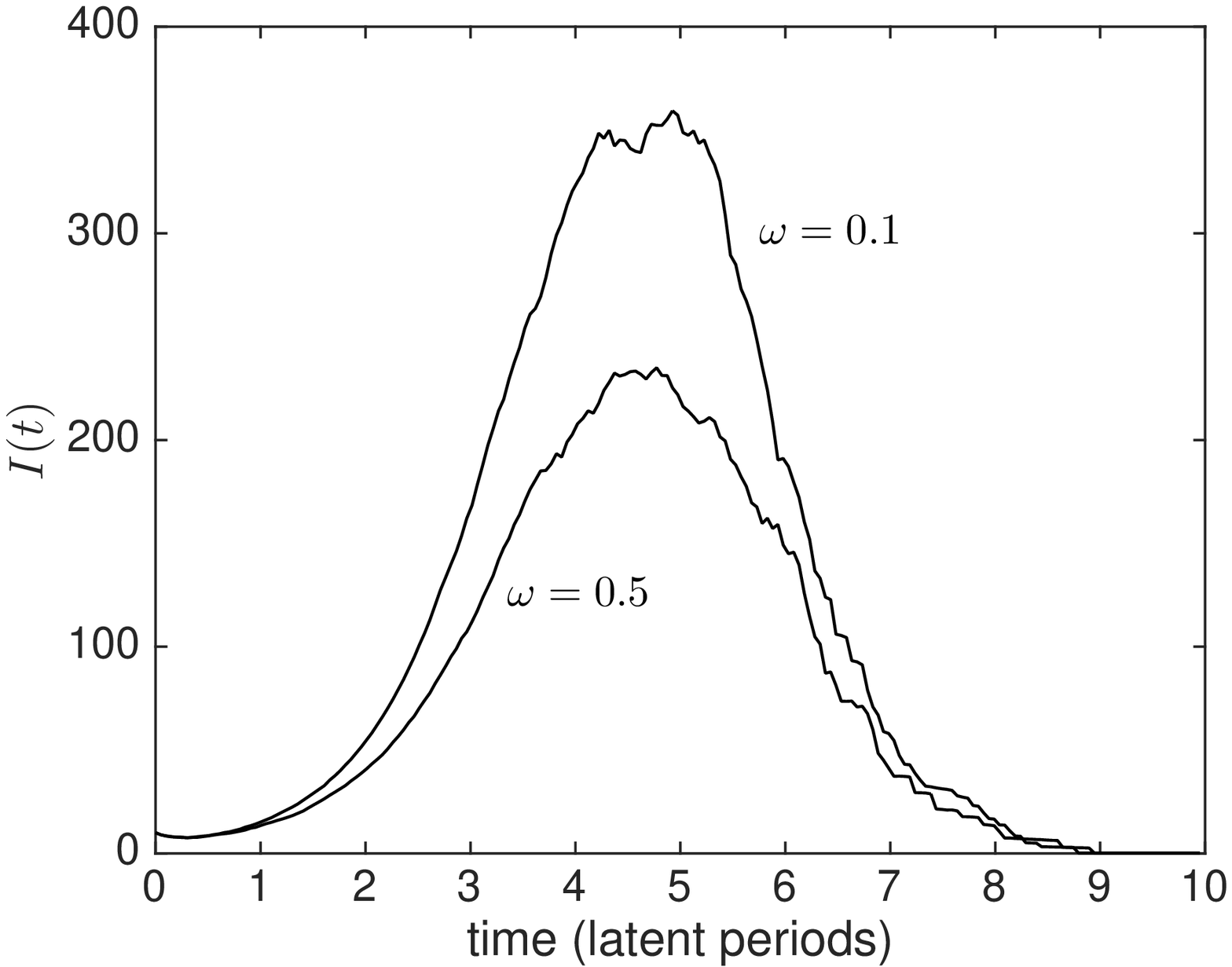}
\end{tabular}
\caption{Mean degree (left) and total number (right) of infectious nodes in an SEIR-$\omega$ epidemic over a SF network with $p(k) \sim k^{-4}$ and $k_{\min}=5$ which amounts to a mean degree $\mu=3 k_{\min} / 2= 7.5$. Outputs averaged over 250 runs. Dashed horizontal lines correspond to the predicted $E(D_I)$: $9.75$ for $\omega=0.1$, and $8.84$ for $\omega=0.5$ (both approaches lead to these rounded values). Parameters: $\phi=1$, $\gamma=2$, $\beta=1.2$, $\alpha=1$, $\omega_{SE}=0$, and $\omega_{SI}=\omega_{EI}=\omega\in\{0.1, 0.5\}$. $N=10000$ and the initial fraction of infected nodes is 0.1\%. \label{SF-SEIR}}
\end{figure*}

As for the SEIR-$\omega$ model, the values of $E(D_I)$ computed from both approaches are very close to each other (see Fig.~\ref{Comparison_E(Di)}) and show a very good agreement with the simulations on Poisson networks (see Fig.~\ref{Poisson-SEIR}). However, the corresponding expressions are not easily manageable (both depend on the solution of a cubic equation when $\alpha \, \omega_{EI} > 0$) and, therefore, they have been evaluated numerically. From these evaluations, it follows that, when $\alpha>0$ and $\omega_{EI}=0$ or, alternatively, when $\alpha=0$ (dropping of edges),  both approaches lead to the same values of $E(D_I)$. For $\alpha \, \omega_{EI} > 0$, predictions are  almost graphically indistinguishable from each other (for $\alpha=1$ and $\omega_{SI}=\omega_{EI}$, the maximum differences occur at second decimal place of the expected degree). As with the SIR-$\omega$ model, when the simulations take place on scale-free networks, the predicted $E(D_I)$ overestimates the observed $\overline{D_I}(t)$ (see Fig.~\ref{SF-SEIR}).

Finally, it is important to note that, for dynamic heterogeneous networks whose degree distribution evolves in time, the value of $R_0$ does not determine the final epidemic size. While the computation of $R_0$ is based on the initial degree distribution, the final epidemic size depends on the whole evolution of the degree distribution. In particular, since reconnection is assumed to be uniform with respect to the degree of nodes, the variance of the degree distribution decreases over time whenever the initial network is highly heterogeneous (see \cite{JRS}). Determining an expression for the final epidemic size, using any approximation method, and studying how it depends on model parameters, remains a highly interesting open problem.

\section*{Acknowledgements}

T.B. is grateful to Riksbankens Jubileumsfond (grant P12-0705:1) for financial support. The research of D.J. and J.S. have been partially supported by the research grants MTM2011-27739-C04-03 and MTM2014-52402-C3-3-P of the Spanish government (MINECO). J.S. and D.J. are members of the research groups 2014SGR-1083 and 2014SGR-555 of the Generalitat de Catalunya, respectively.

\section*{Appendix}

\subsection*{The limit system for the SEIR-$\omega$ model and its equilibria}

The limit system for the local densities involved in the computation of $E(D_I)$ for the SEIR pairwise model when $2[SS]/[S] \to 2L/N$ (the mean degree),
$[SE]/[S] \to 0$, $[SI]/[S] \to 0$, and $[SR]/[S] \to 0$, i.e., at the beginning of an epidemic, is

\begin{align*}
\frac{d}{dt} \left( \frac{[SI]}{[I]} \right) & = - \left( \beta + \omega_{SI} + \phi \frac{[E]}{[I]} \right) \frac{[SI]}{[I]} + \phi \frac{[SE]}{[I]}
\\
\frac{d}{dt} \left( \frac{[SE]}{[I]} \right) & = \beta (\bar{q} - 1) \frac{[SI]}{[I]} - \left( \phi \left( 1 +  \frac{[E]}{[I]} \right) + \omega_{SE} - \gamma  \right) \frac{[SE]}{[I]} + \alpha \, \omega_{EI} \frac{[EI]}{[I]}
\\
\frac{d}{dt} \left( \frac{[E]}{[I]} \right) & = \beta \frac{[SI]}{[I]}  - \left( \phi \left( 1 + \frac{[E]}{[I]} \right) - \gamma \right) \frac{[E]}{[I]}
\\
\frac{d}{dt} \left( \frac{[EI]}{[I]} \right) & = \beta \frac{[SI]}{[I]} - \left( \phi \left( 1+ \frac{[E]}{[I]} \right) + \omega_{EI} \right)  \frac{[EI]}{[I]} + \phi \frac{2[EE]}{[I]}
\\
\frac{d}{dt} \left( \frac{2[II]}{[I]} \right) & = 2 \phi \frac{[EI]}{[I]}  - \left( \phi \frac{[E]}{[I]} + \gamma \right)  \frac{2[II]}{[I]}
\\
\frac{d}{dt} \left( \frac{[IR]}{[I]} \right) & = \phi \frac{[ER]}{[I]} + \gamma \frac{2[II]}{[I]} - \phi \frac{[E]}{[I]} \frac{[IR]}{[I]}
\\
\frac{d}{dt} \left( \frac{2[EE]}{[I]} \right) & = - \left( \phi \left( 2 + \frac{[E]}{[I]} \right) - \gamma \right)  \frac{2[EE]}{[I]}
\\
\frac{d}{dt} \left( \frac{[ER]}{[I]} \right) & = - \left( \phi  \left( 1+ \frac{[E]}{[I]} \right) - \gamma \right) \frac{[ER]}{[I]} + \gamma \frac{[EI]}{[I]}
\end{align*}
where $\bar{q} := E(\tilde{D})$. Note that only one positive term contains a rewiring rate, namely, the last one in the second equation which has $\alpha \, \omega_{EI}$ as a prefactor. Since $[SE]/[I]$ increases with $\alpha \, \omega_{EI}$ and, in turn, the first equation tells us that $[SI]/[I]$ increases with $[SE]/[I]$, the presence of such a positive term means that also $R_0$ increases with $\alpha \, \omega_{EI}$. On the other hand, since the first three equations of the limit system are decoupled from the other five when $\alpha=0$, the same three equations will also govern the dynamics of $[SI]/[I]$ when individuals simply drop connections to infectious neighbours ($\alpha=0$). In particular, this reflects the (expected) fact that whether exposed individuals break off with infectious neighbors ($\omega_{EI}>0$) or not ($\omega_{EI}=0$) does not affect the early dynamics of $[SI]/[I]$ as long as they do not replace these links with new connections.

For a given value $\xi$ of $([E]/[I])^*$, the equilibrium equations (i.e., the equations obtained by making the right-hand side (rhs) of the previous system equal to 0) define a linear system for the remaining variables and, hence, the equilibrium can be easily expressed in terms of $\xi$. A simple inspection of the equations shows that there are two equilibria, $P_1=(0,0,0,0,0,0,0,0)$ and $P_2=(0,0,\gamma/\phi - 1,0,0,0,0,0)$, where the local densities around infectious nodes, $([SI]/[I])^*$, $(2[II]/[I])^*$, $([EI]/[I])^*$, and $([IR]/[I])^*$, are zero. Moreover, from the third equation we see that there exists an equilibrium with $([SI]/[I])^* > 0$ if and only if $([E]/[I])^* > \gamma/\phi - 1$, which in turn implies that $(2[EE]/[I])^*=0$. Consequently, the first four equations at equilibrium do not depend on the last four.

Expressing $([SI]/[I])^*$, $([SE]/[I])^*$, and $([EI]/[I])^*$ in terms of $\xi$, and replacing them into the second equation, it follows that there will be an equilibrium with positive local densities around infectious nodes if there exists a solution $\xi^* > 0$  of the equation
\begin{equation}
(\bar{q} -1) (1+\xi) \phi + (\bar{q}+\alpha-1) \, \omega_{EI} = \frac{\phi^2}{\beta} \left(\xi + \frac{ \phi + \omega_{SE} - \gamma}{\phi} \right) \left (\xi + \frac{\omega_{SI} + \beta}{\phi} \right)  \left( \xi + \frac{\phi + \omega_{EI}}{\phi} \right)
\label{xi}
\end{equation}
such that $\xi^* > \gamma/\phi -1$ if $\gamma > \phi$. Note that, for $\omega_{EI} = 0$, Eq.~\eqref{xi} becomes a quadratic equation in $\xi$ with a positive solution $\xi^*$ given by (see \cite{LJS})
\begin{equation}
\label{xi*}
\xi^* = \frac{1}{2\phi}\left( \gamma - \beta - \phi - \omega_{SI} - \omega_{SE} + \sqrt{(\gamma + \beta - \phi + \omega_{SI} - \omega_{SE})^2 + 4 \phi \beta (\bar{q} -1)} \right)
\end{equation}
with the proviso that $\phi \beta (\bar{q} -1) > (\phi + \omega_{SE} - \gamma)(\beta+\omega_{SI})$. Moreover, in this case ($\omega_{EI}=0$), when $\gamma > \phi$ it follows that $\xi^* > \gamma/\phi -1$ if $\phi \beta (\bar{q} -1) > \omega_{SE} (\gamma + \beta + \omega_{SI} - \phi)$.

If $\gamma \ge \phi$ and $\omega_{EI}>0$, a necessary and sufficient condition for the existence of a unique solution $\xi^*$ of \eqref{xi} such that $\xi^* > \gamma/\phi -1$ is that the straight line defined by the left-hand side (lhs) of \eqref{xi} intersects the vertical line $\xi=\gamma/\phi -1$ above the intersection with this vertical line of the cubic polynomial defined by the rhs of \eqref{xi}. This polynomial has two negative roots, $\xi_1$ and $\xi_2$, and the third root $\xi_3$ can be positive or negative depending on the sign of $\gamma - \phi - \omega_{SE}$. The resulting condition on the parameters is
$$
\phi \beta ( (\bar{q} - 1) \gamma + (\bar{q}+\alpha-1) \, \omega_{EI}) > \omega_{SE} (\gamma + \beta + \omega_{SI} - \phi)(\gamma + \omega_{EI}).
$$
Note that this condition is fulfilled when $\omega_{SE}=0$ which was, indeed, what we assumed in the model if $\omega_{EI}>0$.

If $\gamma < \phi$, then $\xi_3 < 0$ and a sufficient condition for the existence of a $\xi^* > 0$ is that the intersection of the lhs of \eqref{xi} with the $y$-axis is above the intersection with this axis of the rhs of \eqref{xi}. This condition leads to
$$
\phi \beta ( (\bar{q} - 1) \phi + (\bar{q}+\alpha-1) \, \omega_{EI}) > (\phi - \gamma + \omega_{SE}) (\omega_{SI} + \beta) (\phi + \omega_{EI}).
$$
Since we are assuming $\omega_{EI}>0$, then $\omega_{SE}=0$, and from this inequality it follows a simpler sufficient condition for the existence of $\xi^*>0$, namely, $(\bar{q}-2) \beta > \omega_{SI}$.

\subsection*{Relationship between the basic reproduction numbers $R^{BA}_0$ and $R_0^{PA}$ for the SEIR-$\omega$ model}

From \eqref{R0-SEIR-PA}, it follows that the only positive value of $\xi^*$ for which $R_0^{PA} = 1$ is $\xi^* = \gamma/\phi$. After replacing $\xi^*$ by this value, expression \eqref{xi} can be rewritten as
$$
\frac{\phi \beta}{(\phi + \omega_{SE})(\gamma+\beta+\omega_{SI})}
\left( E(\tilde{D}) -1 + \frac{\alpha \, \omega_{EI}}{\gamma+\phi+\omega_{EI}} \right) =1.
$$
Comparing this expression and that of $R_0$ given by \eqref{R0-SEIR}, it follows that $R_0^{PA} = 1 \Leftrightarrow R^{BA}_0 = 1$ if $\alpha \, \omega_{EI}=0$, and that $R_0^{PA} = 1 \Rightarrow R^{BA}_0 > 1$ (and $R^{BA}_0 = 1 \Rightarrow R_0^{PA} < 1$) if $\alpha \, \omega_{EI}>0$. So, for $\alpha >0$, both approximations lead to the same epidemic threshold when $\omega_{EI}=0$.


\section*{References}

\begin{thebibliography}{10}

\bibitem{Anderson}
Anderson, R.M., May, R.M., \ 1991 Infectious diseases of humans: dynamics and control.
Oxford University Press, New York.

\bibitem{BD95}
Ball F. G., Donnelly, P., \ 1995. Strong approximations for epidemic models. Stoch. Proc. Appl. 55, 1-21.


\bibitem{BJM-L}
Britton, T., Janson, S., Martin-L{\"o}f, A., \ 2007. Graphs with specified degree distributions, simple epidemics and local vaccination strategies. Adv. Appl. Prob. 39, 922-948.

\bibitem{DHB}
Diekmann,  O., Heesterbeek, J.A.P., Britton, T., \ 2013. Mathematical tools for understanding infectious diseases dynamics,
Princeton University Press, Princeton.

\bibitem{DHM}
Diekmann, O., Heesterbeek, J. A. P., Metz, J.A.J., \ 1990. On the definition and the computation of the basic reproduction ratio $R_0$ in models for infectious diseases in heterogeneous populations. J. Math. Biol. 28, 365--382.

\bibitem{Durrett}
Durrett, R., \ 2007.  Random Graph Dynamics. Cambridge Series in Statistical and Probabilistic Mathematics. Cambridge University Press, Cambridge.

\bibitem{Eames}
Eames, K.T.D., Keeling, M.J., \ 2002. Modeling dynamic and network heterogeneities in the
spread of sexually transmitted diseases. Proc. Natl. Acad. Sci. USA 99, 13330--13335.

\bibitem{Frasca} Frasca, M., Sharkey, K.J., \ 2016. Discrete-time moment closure models for epidemic spreading in populations of interacting individuals. J. Theor. Biol. 399, 13--21.

\bibitem{Fenichel}
Fenichel, E.P., Castillo-Chavez, C., Ceddia, M.G., et al., \ 2011.
Adaptive human behavior in epidemiological models. Proc. Natl. Acad. Sci. USA 108, 6306--6311.

\bibitem{Gillespie}
Gillespie, D. T., 2007.  Stochastic simulation of chemical kinetics. Annu. Rev. Phys. Chem. 58, 35--55.

\bibitem{Gross06}
Gross, T., DiLima, C.J.D., Blasius, B., \ 2006. Epidemic Dynamics on an Adaptive Network, Phys. Rev. Lett. 96, 208701.

\bibitem{House}
House, T., Keeling, M., \ 2011. Insights from unifying modern approximations to infections on networks. J. R. Soc. Interface 8, 67--73.

\bibitem{Jagers}
Jagers, P., \ 1975. Branching Processes with Biological Applications. Wiley, London.

\bibitem{JRS}
Juher, D., Ripoll, J., Salda\~{n}a, J., \ 2013. Outbreak analysis of an SIS epidemic model with rewiring. J. Math. Biol. 67, 411--432.

\bibitem{Keeling99}
Keeling, M.J., \ 1999. The effects of local spatial structure on epidemiological invasions, Proc. R. Soc. London B 266, 859--867.

\bibitem{KG}
Keeling, M.J., Grenfell, B.T., \  2000. Individual-based perspectives on $R_0$. J. Theor. Biol. 203, 51--61.

\bibitem{Kiss2012}
Kiss, I.Z., Berthouze, L., Taylor, T.J., Simon, P.L. \ 2012. Modelling approaches for simple dynamic networks and applications to disease transmission models. Proc. R. Soc. A 468, 1332--1355.

\bibitem{Lau}
Lau, J.T.F., Griffiths, S., Choi,  K.C., Tsui, H.Y., \ 2010. Avoidance behaviors and negative psychological responses in the general population
in the initial stage of the H1N1 pandemic in Hong Kong. \textit{BMC Infectious Diseases} 10, 139. doi:10.1186/1471-2334-10-139

\bibitem{Lindquist}
Lindquist, J., Ma, J., van den Driessche, P., Willeboordse, F.H., \ 2011. Effective degree network disease models. J. Math. Biol 62, 143--164.

\bibitem{Lipsitch}
Lipsitch, M. et al., \ 2003. Transmission dynamics and control of severe acute respiratory syndrome, Science 300, 1966--1970.

\bibitem{LJS}
Llensa, C., Juher, D., Salda\~{n}a, J., \ 2014. On the early epidemic dynamics for pairwise models. J. Theor. Biol. 352, 71--81.

\bibitem{Marceau}
Marceau, V., No\"{e}l, P.A., H\'ebert-Dufresne, L., Allard, A.,  Dub\'e, L.J., \ 2010. Adaptive networks: Coevolution of disease and topology. Phys. Rev. E 82, 036116.

\bibitem{MHFM}
McCarthy, M., Haddow, L.J., Furner, V., Mindel, A., \ 2007. Contact tracing for sexually transmitted infections in New South Wales, Australia, Sexual Health 4, 21--25.

\bibitem{Miller}
Miller, J.C., \ 2012. A note on the derivation of epidemic final sizes. Bull. Math. Biol. 74, 2125--2141.

\bibitem{MSV}
Miller, J.C., Slim, A.C., Volz, E.M., \ 2012. Edge-based compartmental modelling for infectious disease spread. J. R. Soc. Interface 9, 890--906.

\bibitem{PBT}
Pellis, L., Ball, F., Trapman, P., \ 2012. Reproduction numbers for epidemic models with households and other social structures. Definition and calculation of $R_0$. Math. Biosci. 235, 85--97.

\bibitem{Riley}
Riley, S., et al., \ 2003. Transmission dynamics of the etiological agent of SARS in Hong Kong: Impact of public health interventions, Science 300, 1961--1966.

\bibitem{Risau}
Risau-Gusman, S., Zanette, D.H., \ 2009. Contact switching as a control strategy for epidemic outbreaks. J. Theor. Biol. 257, 52--60.

\bibitem{Schwartz}
Schwartz, I.B., Shaw, L.B., \ 2010. Rewiring for adaptation, Physics 3, 17.

\bibitem{Schwarzkopf}
Schwarzkopf, Y., R\'akos, A., Mukamel, D., \ 2010. Epidemic spreading in evolving networks. Phys. Rev. E 82, 036112.

\bibitem{Sharkey} Sharkey, K.J., \ 2008. Deterministic epidemiological models at the individual level. J. Math. Biol. 57, 311--331.

\bibitem{Spring}
Springborn, M., Chowell, G., MacLachlan, M., Fenichel, E.P., \ 2015.  Accounting for behavioral responses during a flu epidemic using home television viewing. BMC Infectious Diseases 15:21. DOI 10.1186/s12879-014-0691-0

\bibitem{TTK}
Taylor, M., Taylor, T.J., Kiss, I.Z., \ 2012. Epidemic threshold and control in a dynamic network. Phys. Rev. E 85, 016103.

\bibitem{Volz07}
Volz, E.,  Meyers, L.A., \ 2007. Susceptible-infected-recovered epidemics in dynamic contact networks. Proc. R. Soc. B 274, 2925--2933.

\bibitem{Volz09}
Volz, E., Meyers, L.A. \ 2009. Epidemic thresholds in dynamic contact networks. J. R. Soc. Interface 6, 233--241.

\bibitem{Zanette}
Zanette, D.H., Risau-Gusm\'an, S., \ 2008. Infection spreading in a population with evolving contacts. J. Biol. Phys. 34, 135--148.

\end{thebibliography}
\end{document}